\newcommand{\be}{\begin{equation}}\newcommand{\ee}{\end{equation}}
\newcommand{\bea}{\begin{eqnarray}}\newcommand{\eea}{\end{eqnarray}}
\newcommand{\nn}{\nonumber\\[6pt]}
\newcommand{\p}[1]{(\ref{#1})}

\newcommand{\bD}{\overline D}

\newcommand{\bt}{{\bar\theta}}

\newcommand{\beps}{{\bar\epsilon}}

\newcommand{\sfrac}[2]{{\textstyle\frac{#1}{#2}}}

\documentclass[12pt]{article}
\usepackage{amscd,amsmath,amssymb}

\begin{document}

\thispagestyle{empty}
\vspace{2cm}
\begin{flushright}
hep-th/0307111 \\
ITP--UH--05/03 \\[5mm]
July, 2003\\[2cm]
\end{flushright}
\begin{center}
{\Large\bf N=4 Supersymmetric Mechanics \vspace{0.3cm}

in Harmonic Superspace}
\end{center}
\vspace{1cm}

\begin{center}
{\large\bf  E. Ivanov${}^{a}\;$  and  ${}$ O. Lechtenfeld${}^{b}$ }
\end{center}

\begin{center}
${}^a$ {\it Bogoliubov  Laboratory of Theoretical Physics, JINR, 141980 Dubna,
Russia;} {\tt eivanov@thsun1.jinr.ru}

\vspace{0.2cm}
${}^b$ {\it Institut f\"ur Theoretische Physik, Universit\"at Hannover,} \\
{\it Appelstra\ss{}e 2, 30167 Hannover, Germany; }\\
{\tt lechtenf@itp.uni-hannover.de}
\end{center}
\vspace{1cm}

\begin{abstract}
\noindent
We define $N{=}4, d{=}1$ harmonic superspace ${\bf HR}^{1+2|4}$ with an SU(2)/U(1) harmonic part,
SU(2) being one of two factors of the R-symmetry group SU(2)$\times$ SU(2)
of $N{=}4, d{=}1$ Poincar\'e supersymmetry. We reformulate, in this new setting, the
models of $N{=}4$ supersymmetric quantum mechanics associated with the off-shell
multiplets $({\bf 3, 4, 1})$ and $({\bf 4, 4, 0})$. The latter admit
a natural description as constrained superfields living in an analytic subspace of ${\bf HR}^{1+2|4}$.
We construct the relevant superfield actions
consisting of a sigma-model as well as a superpotential parts and demonstrate that
the superpotentials can be written off shell in a manifestly
$N{=}4$ supersymmetric form only in the analytic superspace. The constraints implied
by $N{=}4$ supersymmetry
for the component bosonic target-space metrics, scalar potentials and background one-forms
automatically follow from the harmonic superspace description. The analytic superspace is shown
to be closed under the most general $N{=}4, d{=}1$ superconformal group $D(2,1;\alpha)$. We
give its action on the analytic superfields comprising the
$({\bf 3, 4, 1})$ and $({\bf 4, 4, 0})$ multiplets, reveal a surprising relation
between the latter and present the corresponding superconformally invariant actions.
The harmonic superspace approach suggests a natural generalization of
these multiplets, with a $[{\bf 2(n{+}1), 4n, 2(n{-}1)}]$ off-shell content for $n{>}2$.
\end{abstract}

\newpage
\setcounter{page}{1}
\section{Introduction}
Models of supersymmetric quantum mechanics (SQM) with extended $N{\geq}2, d{=}1$
supersymmetry have plenty of uses (see a recent review \cite{review}).
For instance, they describe the low-energy dynamics of monopoles in $N{=}4$
supersymmetric Yang-Mills theory \cite{gauntl1}. Some variants of
$N{=}4$ superconformal mechanics \cite{nscm} play the role of conformal field
theories in the AdS$_2$/CFT$_1$ correspondence and describe the near-horizon
dynamics of black-hole solutions of supergravity. Supersymmetric extensions of
integrable $d{=}1$ models, e.g. Calogero-Moser type systems \cite{will,bgk},
are expected to have interesting implications in string theory \cite{6}.
It is peculiar that not all $d{=}1$ supersymmetric models can be directly
recovered from appropriate $d{>}1$ theories via dimensional reduction.
They reveal some special target-space geometries which have no direct
counterparts in higher dimensions (see e.g. \cite{papa2,hull}).
For $N{=}4, d{=}1$ models the just-mentioned peculiarity manifests itself, in particular,
in the fact that the most general $N{=}4, d{=}1$ superconformal symmetry is
provided by the supergroup $D(2,1;\alpha)$ \cite{papa1,stro1,stro2} which only for
special values of the parameter $\alpha$ is isomorphic to SU($1,1|2$) obtainable
from higher-dimensional superconformal groups by dimensional reduction.

In many studies of $d{=}1$ supersymmetry (see e.g. \cite{papa2,papa1})
the $d{=}1$ actions invariant under extended supersymmetries are constructed in
components or/and in $N{=}1$ superfields, proceeding from the most general
$N{=}1$ supersymmetric form of such actions and revealing the restrictions which are
imposed on the relevant target geometries by the requirement of invariance with respect to
additional supersymmetries. In such formulations the higher supersymmetries
are non-manifest and frequently on-shell. Like for the case of supersymmetry
in $d{>}1$, it is desirable to have formulations of $d{=}1$ supersymmetric
theories in the appropriate superspaces where {\it all} their underlying
supersymmetries are off-shell and manifest. Then the constraints which ensure
the relevant target-space geometries to be consistent with extended
supersymmetry are valid \'a priori and, in fact, can be read off by studying
the component structure of the action.
For $N{=}4$ supersymmetric mechanics such formulations have been pioneered in
\cite{ikl2,ismi,ikp} and further elaborated e.g. in
\cite{stro2}, \cite{7}--\cite{ikl}. Until present, only the standard type of
$N{=}4, d{=}1$ superspaces was utilized, namely the real $(1|4)$-dimensional
and chiral $(1|2)$-dimensional superspaces ${\bf R}^{1|4}$ and ${\bf C}^{1|2}$.
On the other hand, it is known that many remarkable geometric features of
extended supersymmetric theories are manifest only in harmonic superspace
formulations~\cite{gikos,book}.  One can expect that such formulations visualize
the non-standard target-space geometries of $d{=}1$ models like they visualize
the hyper-K\"ahler and quaternion-K\"ahler geometries of $N{=}2, d{=}4$
supersymmetric sigma models \cite{book} (and their dimensionally-reduced
descendants). Also, a reformulation of $N{=}4$ SQM models in harmonic
superspace might help in constructing SQM models with more than four supersymmetries
by joining models associated with different $N{=}4, d{=}1$ supermultiplets.

As a step toward these goals, in this paper we present the harmonic superspace
formulation of the $N{=}4$ supersymmetric mechanics model proposed and studied in
\cite{cromrit,andr1,andr2,ismi,bepa,stro2} and further elaborated for the $N{=}4$
superconformally
invariant case in \cite{ikl}. It is associated with the off-shell supermultiplet
$({\bf 3, 4, 1})$ comprising three physical and one auxiliary
bosonic fields and four fermionic fields. Furthermore, we discuss along similar
lines a model based on a
different $N{=}4, d{=}1$ supermultiplet which also admits a natural
description in harmonic superspace. The off-shell content of this
supermultiplet is $({\bf 4, 4, 0})$.

The $N{=}4, d{=}1$ harmonic superspace ${\bf HR}^{1+2|4}$ contains in its
bosonic sector two extra harmonic coordinates representing a sphere
$S^2 \sim $ SU(2)/U(1) where SU(2) is one of two commuting SU(2) factors
comprising the full R-symmetry group of $N{=}4, d{=}1$ Poincar\'e
supersymmetry \cite{ikp}.\footnote{
In principle, one can `harmonize' both SU(2) factors and consider a
bi-harmonic superspace of the type employed e.g. in \cite{IvSu} and \cite{zu}.
We limit our study here to the simplest case of just one set of SU(2)
harmonic variables.}
It was introduced in \cite{di} (see also \cite{dik}) in order to construct an
$N{=}4$ superextension of the KdV hierarchy, but was never utilized for $d{=}1$
sigma model building.

The paper is organized as follows. In Sections 2 and 3 we recollect the necessary
facts about the standard $N{=}4, d{=}1$ superspace and its SU(2) harmonic extension.
Then in Sections 4 and 5 we present the harmonic superspace formulation
of $N{=}4$ SQM models
associated with the $({\bf 3,4,1})$ supermultiplet, with the emphasis
on the superconformally invariant model considered  in \cite{ikl}.
We show that the kinetic and potential terms of its superfield action admit a transparent
presentation in harmonic superspace. In particular, the $N{=}4$
superconformally invariant potential term can be written as an integral over the
$(1|2)$-dimensional analytic subspace ${\bf AR}^{1+2|2}\subset{\bf HR}^{1+2|4}$,
and it is the simplest $d{=}1$ analog of the $d{=}4$ harmonic superspace action
for the {\it improved $N{=}2$ tensor multiplet\/} \cite{gio1,gio2}. We also
present the general off-shell form of an $N{=}4$ supersymmetric (but generically not
conformally invariant) superpotential for this type of $N{=}4$ SQM models.
It is also given by an integral over the analytic superspace which thus provides
the unique possibility to write the superpotential in a manifestly $N{=}4$ supersymmetric
way. The harmonic superspace representation allows one to easily recognize
the general constraints
which $N{=}4$ supersymmetry imposes on the purely potential term in the
component action and on the related term which describes the coupling to an
external three-dimensional gauge potential. In Section 6 we discuss another
variant of $N{=}4$ SQM associated with a $d{=}1$ analog of the four-dimensional
$N{=}2$ {\it hypermultiplet\/}. It carries four
physical bosons and four physical fermions off-shell and has no auxiliary
fields at all.\footnote{Beyond harmonic superspace, such an off-shell
$N{=}4, d{=}1$ multiplet was
discussed e.g. in \cite{stro1,gps,pt,hp,hull}.}It also admits a simple description as the
analytic harmonic $N{=}4, d{=}1$ superfield. We construct superconformally invariant
actions as well as general actions for this multiplet and find the general restrictions
on the relevant
target-space metric and background one-form potential. We also discover an unexpected
relation of this $({\bf 4, 4, 0})$ multiplet with the $({\bf 3, 4, 1})$ multiplet.

\section{Preliminaries: the standard N=4, d=1 superspace}
For further reference, following ref. \cite{ikl}, we quote here some basic relations of
the description of
the $N{=}4$ models of refs. \cite{cromrit,andr1,ismi,bepa,stro2,ikl} in ordinary
$N{=}4$ superspace
\be \label{realR}
{\bf R}^{1|4} = \{ t, \theta_i, \bt^k \} \equiv \{ z \}\,.
\ee

The standard $N{=}4, d{=}1$ Poincar\'e supersymmetry and special conformal supersymmetry
from the most general $N{=}4, d{=}1$ superconformal group $D(2,1;\alpha)$ are realized
on these coordinates by the following transformations,
\bea \label{Poinc}
&& \delta t =
i\left(\theta\cdot\bar\varepsilon -\varepsilon\cdot\bar\theta\right)\ ,\quad
\delta \theta_i = \varepsilon_i\, \quad
\delta\bar\theta^i = \bar\varepsilon^i\ ; \\[6pt]
&& \delta' t =
-it \left( \epsilon \cdot\bt +\beps\cdot\theta \right) +(1+2\alpha)
\theta\cdot\bt \left( \epsilon\cdot\bt -\beps\cdot\theta \right)\ ,\nonumber\\
&& \delta' \theta_i =
\epsilon_i t -2i \alpha \theta_i (\theta\cdot \beps) + 2i (1+\alpha) \theta_i
(\bt\cdot\epsilon) -i (1+2\alpha) \epsilon_i (\theta\cdot\bt)\ ,
\label{Sconf}
\eea
where $\varepsilon_i$ and $\epsilon_i$ are the corresponding SU(2) doublet transformation
parameters.\footnote{We use the short-hand notation  $\bar\psi\cdot \xi = \bar\psi^i\xi_i
= -\xi_i\bar\psi^i = - \xi\cdot\bar\psi\,$, $\psi\cdot \xi = \psi^i\xi_i\,$,
$\bar\psi\cdot \bar\xi = \bar\psi_i\bar\xi^i$. The
SU(2) indices are raised and lowered with the help of the skew-symmetric tensors $\epsilon_{ik},
\epsilon^{ik}\; (\epsilon_{12} = -\epsilon^{12} =1)$.}
All other transformations in $D(2,1;\alpha)$ can be obtained by commuting
these basic ones (together with their complex conjugates). So the generic $N{=}4$
supersymmetric models should respect invariance under
\p{Poinc}, while the superconformal models in addition should be invariant under \p{Sconf}
(perhaps only for some special values of the parameter $\alpha$). The conformal
supergroup $D(2,1;\alpha)$ includes as
a subgroup not only the $N{=}4$ Poincar\'e supergroup but also its R-symmetry group,
i.e. SU(2)$\times$ SU(2). Both these SU(2) factors appear in the commutator of two
supersymmetries \p{Poinc}, \p{Sconf}, but in our notation only
one SU(2) is manifest, namely the one rotating doublet indices $i, j$.
The other SU(2) mixes $\theta_i$ and
$\bar\theta_i$. Both SU(2) can be made manifest by passing to the quartet
notation $(\theta_i, \bar\theta_i) \equiv
\theta^{ia}$ \cite{ikp}, with the second SU(2) then acting
on the additional doublet index $a$. Here we shall not use
this notation.

The semi-covariant (fully covariant only under Poincar\'e supersymmetry)
spinor derivatives are defined by
\be \label{semicD}
D^i=\frac{\partial}{\partial\theta_i}+i\bt^i \partial_t\; , \;
\bD_i=\frac{\partial}{\partial\bt^i}+i\theta_i \partial_t\; , \;
\left\{ D^i, \bD_j\right\}= 2i \delta^i_j \partial_t \; .
\ee
They properly transform through each other under \p{Sconf} (see \cite{ikl}).

The measure of integration over ${\bf R}^{1|4}$ is defined as
\be
\mu \equiv dt d^4\theta = \sfrac{1}{4}\, dt\, (D)^2 (\bar D)^2\,, \quad \int d^4\theta\,
(\theta)^2(\bar\theta)^2 = 4\,, \label{measure1}
\ee
where $(D)^2 = D^iD_i\,, \,(\bar D)^2 = \bar D_i \bar D^i\,$, and its specific
normalization is chosen
for further convenience. It is invariant under \p{Poinc} and transforms with a weight 1
(in mass units)
under the superconformal transformations \p{Sconf}
\be
\delta'\; dtd^4\theta = 2i(\epsilon\cdot \bt + \bar\epsilon\cdot \theta)\,dtd^4\theta~.
\label{meastrans}
\ee

The basic $N{=}4, d{=}1$ superfield in the version of $N{=}4$ mechanics \cite{ismi,bepa,ikl}
we are considering here is the isovector superfield
\be
V^{ik}(z) = V^{ki}(z)\, \;\;\overline{V^{ik}} = \epsilon_{ii'}\epsilon_{kk'}V^{i'k'}\,,
\ee
which is subject to the constraints
\be
D^{(i}V^{kl)} = 0\,\quad \bar D^{(i}V^{kl)} = 0\,.\label{constrV1}
\ee
They leave in $V^{ik}$ the off-shell irreducible component field content
$(\bf{3}, \bf{4}, \bf{1})$:
a real triplet of physical bosonic fields, a complex doublet of physical fermionic
fields, and
a singlet auxiliary field. These constraints are consistent with
the action of $D(2,1;\alpha)$
supersymmetry provided $V^{ik}$ is transformed as
\be
\delta' V^{ij} = -2i\alpha\left[ (\epsilon\cdot\bt+\beps\cdot\theta)V^{ij} +
(\epsilon^{(i}\bt_k-\beps_k\theta^{(i})V^{j)k}+
(\epsilon_k\bt^{(i}-\beps^{(i}\theta_k)V^{j)k} \right].\label{Vtrans}
\ee
As a consequence, the object $V^2 \equiv V^{ik}V_{ik}$ is transformed as a density of
the weight $-2\alpha$:
\be
\delta' V^2 = -4i\alpha\,(\epsilon\cdot\bt+\beps\cdot\theta)\,V^2~. \label{Vtrans2}
\ee
Some useful corollaries of the constraints \p{constrV1} are the following,
\bea
&&(D)^2 V^{ik} = (\bar D)^2 V^{ik} = [D, \bar D] V^{ik} =0~, \nn
&& (D)^2 \left( V^2\right)^{-\frac{1}{2}}=(\bD){}^2 \left( V^2\right)^{-\frac{1}{2}} =
[D, \bD{}] \left( V^2\right)^{-\frac{1}{2}}= 0~, \label{prop1} \\
&& D_i \,\frac{ V^{ij} }{ \left( V^2\right)^{\frac{3}{2}} } =
\bD_i\, \frac{ V^{ij} }{ \left( V^2\right)^{\frac{3}{2}} } =0 \;.
\label{prop2}
\eea

The general sigma-model type action of $V^{ik}$ possesses only $N{=}4, d{=}1$
super Poincar\'e invariance and is given by an integral over ${\bf R}^{1|4}$,
\be
S_{(V)} = -\gamma \int dtd^4\theta \,L(V) \label{kin1}\ ,
\ee
where $\gamma$ is a positive normalization constant and $L(V)$ is an arbitrary
function of $V^{ik}$.
In what follows we shall need the bosonic component part of this action.
It can be directly
obtained using \p{measure1}, \p{constrV1} as
\be
\hat{S}_{(V)} = \gamma\, \int dt H(v) \left(\dot{v}^{ik}\dot{v}_{ik}
+ \sfrac{1}{2} F^2 \right)\,,\qquad
H(v) = \Delta L(v)\,,\label{bos1}
\ee
where
\be
\Delta = \frac{\partial^2}{\partial v^{ik}\partial v_{ik}}\,, \qquad v^{ik}(t) = V^{ik}(z)|\,,
\qquad F(t) = \sfrac{i}{3} \bar D_i D_k V^{ik}(z)|
\ee
and $|$ denotes the restriction to the point $\theta = \bar\theta = 0$. We see that
for bosonic physical fields there arises a sigma model on a 3-dimensional conformally flat
manifold with the Weyl factor $H(v) = \Delta L(v)$. In the case of $n$ superfields
$V^{ik}_A\,, \;\; A=1, \ldots n$, and $L(V_1, V_2,\ldots)$ in \p{kin1} one obtains
the following generalization of \p{bos1} \cite{stro2}:
\bea
&& \hat{S}^{\;n}_{(V)} =  \gamma \int dt \left\{H^{AB}(v) \left(\dot{v}_A^{ik}\dot{v}_{B\,ik}
+ \sfrac{1}{2} F_A F_B \right)
+ 2 G^{[AB]}_{(ik)}\left(\dot{v}^i_{A\,j}\dot{v}^{jk}_B
+ F_{[A}^{} \dot{v}^{ik}_{B]}\right)\right\}\label{bosN} \\
&& H^{AB} = \Delta^{AB} L(v_1,v_2,\ldots )\,,\;\Delta^{AB} =
\frac{\partial^2}{\partial v_A^{ik}\partial v_{B\,ik}}\,,
\quad G^{[AB]}_{(ik)} = \epsilon^{mn}\frac{\partial^2 L}{\partial v^{m(i}_A\partial v^{k)n}_B}\,.
\label{defHG}
\eea
The eventual $3n$-dimensional target space metric arises after integrating out the
auxiliary field $F_A$. Its
explicit form is not too illuminating. We only point out that in the sigma model target
space we encounter
a special type of $3n$-dimensional geometry which is a generalization
of the conformally-flat 3-geometry
in the sense that both are fully specified by a scalar real function $L(v)$
or $L(v_1, v_2, \ldots)$.
In some detail this geometry was studied in \cite{stro2}.\footnote{Actually,
it is a geometry with torsion
which appears in fermionic terms \cite{stro2} and it is a generalization of
the so-called weak HKT geometry \cite{gps}.}
Note that for the special case of $L(V_1, V_2,\ldots )= \sum_A L(\tilde{V}_A)$
where $\tilde{V}^{ik}_A$ are
some linear combinations of the original $V^{ik}_A$, the second term
in \p{bosN} disappears and
the target space metric is drastically simplified. Such systems were
considered e.g. in \cite{andrei,bgk}.

As was shown in \cite{ikl}, for one $V^{ik}$ the $N{=}4$ superconformally invariant
models in superspace
correspond to the following specific choice of the function $L(V)$ in \p{kin1},
\bea
&& S_{(V)}^{conf}(\alpha) = -\gamma\, \int dt d^4\theta\, (V^2)^{\frac{1}{2\alpha}}
\qquad
\textrm{for}\quad  \alpha \neq -1\,, \label{neq-1} \\
&& S_{(V)}^{conf}(\alpha) = \sfrac12\gamma\, \int dt d^4\theta\, (V^2)^{-\frac{1}{2}}
\ln (V^2)\qquad
\textrm{for}\quad \alpha = -1\,. \label{eq-1}
\eea
The second invariant corresponds to the special case when $D(2,1;\alpha)$ becomes
isomorphic to a semi-direct product
of SU($1,1|2$) and second R-symmetry SU(2) group. The first invariant exists
in this case too, but in virtue
of the relations \p{prop1} it is identically vanishing.
Both invariants yield similar bosonic lagrangians
calculated by the general formula \p{bos1} (with the auxiliary fields eliminated)
\be
\hat{S}^{conf}_{(V)}(\alpha) = \gamma\, \sfrac{(1+\alpha)}{\alpha^2}
\int dt (v^2)^{\frac{1-2\alpha}{2\alpha}} \dot{v}^{ik}\dot{v}_{ik}\,,
\quad  \hat{S}^{conf}_{(V)}(\alpha=-1) = \gamma\, \int dt (v^2)^{-3/2}
\dot{v}^{ik}\dot{v}_{ik}\,. \label{confbos}
\ee

In \cite{ikl} we have also shown the existence of a non-trivial $N{=}4$ superconformally
invariant superfield potential term
which in the bosonic sector yields a combination of two well known $d{=}1$
conformal invariants:
the standard potential of conformal mechanics \cite{dff} and the coupling of
a non-relativistic particle in ${\bf R}^3$
to the vector potential of a Dirac magnetic monopole \cite{jack}.
This superinvariant was presented in two equivalent forms:
in $N{=}2, d{=}1$ superspace (where its $N{=}4$ supersymmetry is non-manifest)
and in the full $N{=}4$ superspace (in terms of
an unconstrained prepotential solving \p{constrV1}). We shall demonstrate
that this superpotential
admits a nice manifestly $N{=}4$ supersymmetric representation
in the analytic subspace of the $d{=}1$ harmonic superspace
to be defined below. We shall see that this object is a representative
of the whole class of
superpotentials which in general respect only $N{=}4, d{=}1$ Poincar\'e
supersymmetry and naturally `live' in the analytic harmonic superspace.

\setcounter{equation}{0}
\section{N=4, d=1 harmonic superspace}
The $N{=}4, d{=}1$ harmonic superspace ${\bf HR}^{1+2|4}$ \cite{di} is obtained
by adding to the
$N{=}4, d{=}1$ superspace coordinates \p{realR} a set of harmonic coordinates
$u^\pm_i$ parametrizing
the sphere $S^2 \sim $ SU(2)/U(1), with SU(2) being the R-symmetry group
which acts on the doublet indices $i, j$:
\be
u^+_i, u^-_k \in SU(2)\,, \quad u^{+ i}u^-_i = 1\,,
\;\;u^+_iu^-_k - u^+_ku^-_i = \epsilon_{ik}\,.\label{defu}
\ee
Then one can choose the so called analytic basis in
${\bf HR}^{1+2|4} = {\bf R}^{1|4}\times S^2$
\be
{\bf HR}^{1+2|4} =\{t_A, \theta^+, \bar\theta^+, \theta^-,
\bar\theta^-, u^+_i, u^-_k \}
\equiv \{z_A, u^+_i, u^-_k\}
\equiv \{\zeta,  u^+_i, u^-_k, \theta^-, \bar\theta^-\}\, \label{analbas}
\ee
where
\be
\theta^\pm = \theta^iu^\pm_i\,, \;\; \bar\theta^\pm =
\bar\theta^iu^\pm_i \qquad\textrm{and}\quad
t_A = t - i(\theta^+\bar\theta^- + \theta^-\bar\theta^+)\,. \label{centranal}
\ee
The original basis will be referred to as the central basis.
The analytic basis makes manifest the existence
of an important subspace in ${\bf HR}^{1+2|4}$,
the analytic superspace
${\bf AR}^{1+2|2}$ which is a quotient of \p{analbas}
by $\{ \theta^-, \bar\theta^-\}$, i.e.
\be
{\bf AR}^{1+2|2} = \{\zeta, u \} =
\{t_A, \theta^+, \bar\theta^+, u^+_i, u^-_k \}\,.\label{analSS}
\ee
Its basic feature is that it is closed under the action
of $N{=}4, d{=}1$ Poincar\'e supersymmetry \p{Poinc}
(and under the $D(2,1;\alpha)$ transformations, see below),
\be
\delta \theta^+ = \varepsilon^iu^+_i \equiv \varepsilon^+\,, \;
\delta \bar\theta^+ = \bar\varepsilon^iu^+_i \equiv \bar\varepsilon^+\,, \;
\delta t_A = -2i \left(\varepsilon^-\bar\theta^+
+\theta^+\bar\varepsilon^-\right)\,,\;
\delta u^\pm_i = 0\,.
\ee
This property is closely related to the fact that the harmonic
projections of the spinor covariant derivatives \p{semicD},
\be
D^\pm \equiv D^iu^\pm_i\,, \;\; \bar D^\pm \equiv \bar D^iu^\pm_i\,,
\ee
take the following explicit form in the analytic basis:
\be
D^+ = \frac{\partial}{\partial \theta^-}\,, \;\; \bar D^+ =
-\frac{\partial}{\partial \bar\theta^-}\,, \;\;
D^- = -\frac{\partial}{\partial \theta^+} + 2i \bar\theta^-\partial_A\,, \;\;
\bar D^- = \frac{\partial}{\partial \bar\theta^+} + 2i \theta^- \partial_A\,.
\ee
We see that $D^+, \bar D^+$ become partial derivatives. Then, the
covariant irreducibility conditions
for some superfield $\Phi$ given on ${\bf HR}^{1+2|4}\,$,
\be
D^+ \Phi (z, u) = \bar D^+ \Phi(z, u) = 0 \,, \label{analytcond}
\ee
are recognized in the analytic basis as Grassmann analyticity conditions.
The latter state that
in this basis $\Phi(z, u)$ is independent of $\theta^-, \bar\theta^-$:
\be
\frac{\partial}{\partial \theta^-}\Phi(z_A, u) =
\frac{\partial}{\partial \bar\theta^-}\Phi(z_A, u) = 0\qquad
\Longrightarrow \qquad
\Phi (z_A, u) = \phi (\zeta ) \ .
\ee
In general, analytic superfields can carry external charges with respect
to the harmonic U(1) (denominator of SU(2)/U(1) $\sim S^2$).
These superfields  are assumed to admit harmonic expansions on $S^2$, running
over integer isospins for even external U(1) charges and over half-integer
ones for odd charges.

An important property of both the harmonic superspace and its analytic subspace
is their reality under
the generalized involution which is the product of ordinary complex conjugation
and Weyl
reflection of $S^2$ (antipodal transformation). Details can be found
in \cite{book};
here we only give the transformations of the coordinates and spinor derivatives
in the analytic basis,
\bea
&&\widetilde{t_A} = t_A\,, \;\; \widetilde{\theta^\pm} = \bar\theta^\pm\,,\;\;
\widetilde{\bar\theta^\pm} = -\theta^\pm\,, \;\;
\widetilde{(u^\pm_i)} = u^{\pm i}\,, \;\;  \widetilde{(u^{\pm i})} =
- u^{\pm}_i\,, \;\;\label{tildethet} \\
&& \widetilde{D^\pm} = -\bar D^\pm\,, \;\; \widetilde{\bar D^\pm} = D^\pm\,.
\eea
Using this involution, one can impose reality conditions on the analytic superfields.
The involution squares to $1$ on the objects with even U(1) charges and to $-1$
on those with odd charges.

Another important ingredient of the harmonic formalism are the covariant
derivatives on the harmonic $S^2$. In the central basis they are
\bea
&& D^{\pm\pm} = u^\pm_i\frac{\partial}{\partial u^\mp_i} \equiv \partial^\pm\,,
\;\; D^0 =
u^+_i \frac{\partial}{\partial u^+_i} - u^-_i \frac{\partial}{\partial u^-_i}
\equiv \partial^0\,, \label{DDc} \\
&& [D^{++}, D^{--}] = D^0\,, \quad [D^0, D^{\pm\pm}] = \pm  2 D^{\pm\pm}\,.
\label{Dalg}
\eea
The same objects in the analytic basis read
\bea
&& D^{++} = \partial^{++} - 2i \theta^+\bar\theta^+\partial_A
+ \theta^+\frac{\partial}{\partial \theta^-} +
\bar\theta^+\frac{\partial}{\partial \bar\theta^-}\,, \label{D++a} \\
&&D^{--} = \partial^{--} - 2i \theta^-\bar\theta^-\partial_A
+ \theta^-\frac{\partial}{\partial \theta^+} +
\bar\theta^-\frac{\partial}{\partial \bar\theta^+}\,,\label{D--a} \\
&&   D^0 = \partial^0 + \left(\theta^+\frac{\partial}{\partial \theta^+}
+ \bar\theta^+\frac{\partial}{\partial \bar\theta^+} \right)
- \left(\theta^-\frac{\partial}{\partial \theta^-} +
\bar\theta^-\frac{\partial}{\partial \bar\theta^-} \right)\;.\label{D0}
\eea
These operators are invariant under the $\widetilde{\;\;}$ conjugation.
As is seen from its explicit form,
the covariant derivative $D^{++}$ commutes with the spinor derivatives
$D^+, \bar D^+$ in \p{analytcond} and
so preserves the Grassmann harmonic analyticity: acting on an analytic superfield,
it produces an analytic
superfield. In contrast, the derivative $D^{--}$ does not preserve the analyticity.
The operator $D^0$ counts the
external U(1) charges of the harmonic superfields. Some important
(anti)commutation relations to be
used below are
\be
[D^{\pm\pm}, D^{\mp}] = D^{\pm}\,, \quad [D^{\pm\pm}, \bar D^{\mp}] =
\bar D^{\pm} \quad\textrm{and}\quad
\{ D^+, \bar D^- \} = - \{D^-, \bar D^+\} = 2i\,\partial_t\,. \label{Alg1}
\ee

Now let us see how the superconformal $D(2,1;\alpha)$ symmetry acts
in the harmonic superspace. We start with the ansatz
\be
\delta' u^+_i = \Lambda^{++}u^-_i \qquad\textrm{and}\qquad
\delta' u^-_i =0\,, \label{Stranharm}
\ee
which is consistent with the defining condition $u^{+i}u^-_i = 1$
and is the typical superconformal
transformation law of harmonic variables \cite{book}. Though this is not consistent
with ordinary
complex conjugation, it nicely matches with the above-mentioned generalized
conjugation~$\widetilde{\;\;}$
which substitutes the ordinary one in harmonic superspace. Then we require
the analytic subspace \p{analSS} to be closed under the transformations \p{Sconf}
and \p{Stranharm}
with taking into account the relations \p{centranal}. This requirement proves
to uniquely fix the
transformations of $\theta^+, \bar\theta^+$ and the function $\Lambda^{++}$ as
\bea
&&\delta' \theta^+ = \epsilon^+\,t_A + 2i(1{+}\alpha)
\epsilon^-(\theta^+\bar\theta^+)\,, \quad
\delta' \bar\theta^+ = \bar\epsilon^+\,t_A +
2i(1{+}\alpha)\bar\epsilon^-(\theta^+\bar\theta^+)\,,
\label{trthet+} \\[6pt]
&&\Lambda^{++} = -2i\alpha(\epsilon^+ \bar\theta^+ {-} \bar\epsilon^+ \theta^+ )
\equiv D^{++}\Lambda\,, \quad
\Lambda = -2i\alpha(\epsilon^- \bar\theta^+ {-} \bar\epsilon^- \theta^+ )\;.
\label{defLambda}
\eea
so that $(D^{++})^2\Lambda = 0$.
It is easy to find the transformation laws of $\theta^-, \bar\theta^-$ and $t_A$:
\bea
&& \delta' \theta^- = \epsilon^- t_A + 2i[\,(1{+}\alpha) \theta^+\bar\theta^-
+ \theta^-\bar\theta^+\,]
+ 2i\alpha \,\bar\epsilon^-\theta^-\theta^+
-2i(1{+}\alpha) \epsilon^+ \theta^-\bar\theta^-\,, \label{trthet-} \\[6pt]
&& \delta' t_A = -2i t_A\,(\epsilon^-\bar\theta^+  {-} \bar\epsilon^-\theta^+)\,,
\quad \delta' \bar\theta^- = \widetilde{\delta' \theta^-}\,.
\eea
We see that ${\bf AR}^{1+2|2}$ defined in \p{analSS} is closed under $N{=}4$ superconformal
transformations at any value
of $\alpha $, in contrast to the standard left-chiral subspace of ${\bf R}^{1|4}$, i.e.
${\bf C}^{1|2} = \{t_c, \theta_i \}$,
which is closed only for $\alpha = -1$, i.e. with respect to SU($1,1|2$)
supersymmetry \cite{ikl}. Thus we
can define $N{=}4, d{=}1$ superconformally-covariant {\it analytic\/} harmonic superfields
for any value of $\alpha$, while the analogous
{\it chiral\/} superfields can be consistently defined only for the choice
of SU($1,1|2$) as the $N{=}4, d{=}1$ superconformal
group. It is worth noting that the Grassmann coordinate transformation laws \p{trthet+},
\p{trthet-} are drastically
simplified at $\alpha = -1$.

Now it is easy to find the $D(2,1;\alpha)$ transformation rules of the different
covariant derivatives and
to check e.g. that $D^+, \bar D^+$ in the analytic basis transform through each other,
thereby preserving the Grassmann
analyticity~\p{analytcond}. For our further discussion we shall need the
transformation rules of harmonic
derivatives $D^{++}, D^{--}$ defined in \p{D++a}, \p{D--a},
\be
\delta' D^{++} = -\Lambda^{++}\,D^0 \qquad \textrm{and}\qquad
\delta' D^{--} = -(D^{--}\Lambda^{++}) D^{--}\,.\label{harmDtran}
\ee
The U(1) charge counter $D^0$ is invariant, which can be proved e.g.
by varying the relations of the $D$-algebra \p{Dalg}.
The whole set of $D(2,1;\alpha)$ transformations can be obtained
by repeatedly commuting
the above $\delta'$-transformations with each other and with those of
$N{=}4, d{=}1$ Poincar\'e supersymmetry.
For any element of $D(2,1;\alpha)$,
the transformation of the harmonics and those of $D^{\pm\pm}$ have the same form,
with all transformation parameters
being properly accommodated by the superfunction $\Lambda^{++}(\zeta, u)$.
The latter satisfies the same differential
constraints as in the particular case \p{defLambda}, namely
\be
\Lambda^{++} = D^{++}\Lambda \qquad\textrm{and}\qquad
D^{++}\Lambda^{++} = 0\,.\label{genLambd}
\ee

The measures of integration over the full harmonic superspace and over
its analytic subspace, denoted by $\mu_H$ and $\mu_A$,
are defined as
\bea
&& \mu_H = du\,\mu = du dt d^4\theta = du dt_A (D^-\bar D^-)( D^+\bar D^+) =
\mu_A^{--} (D^+\bar D^+)\;,\nn
&& \mu_A^{--} = du d\zeta^{--} = du dt_A D^-\bar D^-\,. \label{measHA}
\eea
They are evidently invariant under the $N{=}4$ Poincar\'e supersymmetry and
have the following
transformation properties under the superconformal transformations:
\bea
\delta' \mu_A^{--} &=&
\left(\partial_A\delta' t_A + \partial^{--}\Lambda^{++} -
\partial_{\theta^+}\delta' \theta^+ -
\partial_{\bar\theta^+}\delta' \bar\theta^+\right)\mu_A^{--} = 0\,,
\label{muA}\\[6pt]
\delta' \mu_H &=&
\left(\partial_A\delta' t_A + \partial^{--}\Lambda^{++} -
\partial_{\theta^+}\delta' \theta^+ -
\partial_{\bar\theta^+}\delta' \bar\theta^+
- \partial_{\theta^-}\delta' \theta^- -
\partial_{\bar\theta^-}\delta' \bar\theta^-\right)\mu_H \nonumber\\
&=& 2i\left[ (1{-}\alpha)(\epsilon^-\bar\theta^+ -\bar\epsilon^-\theta^+) -
(1{+}\alpha)(\epsilon^+\bar\theta^- -\bar\epsilon^+\theta^-)\right]\mu_H\,.
\label{muH}
\eea
Recall that the integration over the harmonic $S^2 = \{u^+_i, u^-_k\}$ is
normalized so that
\be
\int du\, 1 = 1\,,
\ee
and the integral of any other irreducible monomial of the harmonics
is vanishing \cite{book}.

Finally, let us point out that the $N{=}4, d{=}1$ harmonic superspace
is by no means a dimensional reduction of
the $N{=}2, d{=}4$ one. Indeed, $N{=}2, d{=}4$ supersymmetry amounts
to $N{=}8$ supersymmetry in $d{=}1$, so
a $d{=}1$ reduction of theories in $N{=}2, d{=}4$ harmonic superspace would
yield $N{=}8$ supersymmetric SQM
models. On the other hand, the $N{=}4, d{=}1$ supersymmetry can be regarded
as a reduction of the $N{=}1, d{=}4$
one. No standard harmonic superspaces can be defined for the latter
because of lacking of non-abelian
R-symmetry group in the $N{=}1, d{=}4$ case. However, such symmetry
(SU(2)$\times$ SU(2)) appears after reduction to
$d{=}1$ and this makes it possible to define harmonic superspace
for $N{=}4, d{=}1$ supersymmetry.

\setcounter{equation}{0}
\section{Constrained analytic N=4, d=1 superfields}
Generically, unconstrained analytic $N{=}4, d{=}1$ superfields contain an
infinite number of standard $d{=}1$ fields with growing isospin, due to
the harmonic $S^2$ expansions of the $u$-dependent component fields
in the $\theta^+, \bar\theta^+$ expansion.
While in higher dimensions the presence of these infinite tails
of auxiliary fields
sometimes turns out to be crucial for the existence of off-shell
superfield actions,\footnote{
Take, for instance, the off-shell actions of
the hypermultiplets in $N{=}2, d=4$ supersymmetry \cite{gikos,book}.}
for the time being it is far from obvious
which role such unconstrained superfields could play in
$N{=}4, d{=}1$ SQM models. On the other hand, one can suitably
constrain the harmonic dependence by imposing on these superfields
harmonic constraints which make use of the
analyticity-preserving harmonic derivative $D^{++}$. In this way not
only some known $N{=}4, d{=}1$ multiplets can be
recovered as constrained $N{=}4, d{=}1$ harmonic analytic superfields,
but also new types of such multiplets
can be exhibited. To be aware of all such multiplets is important
e.g. for $N>4, d{=}1$ model building.

Let us start by showing that the constraints \p{constrV1} are nothing
but the Grassmann harmonic analyticity conditions.
One introduces a harmonic superfield $V^{++}(z, u)$ and subjects
it to the following
set of constraints in harmonic superspace:
\be
(\mbox{a})\quad D^+ V^{++} = \bar D^+ V^{++} = 0\,; \qquad
(\mbox{b})\quad D^{++} V^{++} = 0\,. \label{Hconstr}
\ee
The constraint (\ref{Hconstr}b) in the central basis simply implies $V^{++}$
to be quadratic in the harmonics:
\be
D^{++}V^{++}(z, u) = 0\; \qquad \Longrightarrow \qquad V^{++}(z,u) =
V^{ik}(z) u^+_iu^+_k\;.
\ee
Then (\ref{Hconstr}a), after stripping off the harmonics,
yields just \p{constrV1}. The virtue
of this equivalent form (\ref{Hconstr}) for the basic constraints \p{constrV1}
is revealed after passing to the
analytic basis \p{analbas}. In it, eqs. (\ref{Hconstr}a) are just
the Grassmann analyticity conditions
\bea
V^{++} = V^{++}(\zeta, u)\, \label{Vanal}
\eea
while (\ref{Hconstr}b) cuts the infinite harmonic expansion of $V^{++}(\zeta, u)$
to finite size, leaving in it just the irreducible $({\bf 3,4,1})$ field content
\bea
V^{++}(\zeta, u) = v^{ik}(t_A)u^+_iu^+_k  + \theta^+ \psi^i(t_A)u^+_i
+ \bar\theta^+ \bar\psi^i(t_A)u^+_i + i\theta^+\bar\theta^+\left(F(t_A)
+ 2\dot{v}^{ik}(t_A)u^+_iu^-_k\right),\nonumber
\eea
\be
{\,} \label{solv(b)}
\ee
where $v^{ik}$ and $F$ are the physical and auxiliary bosonic fields, respectively.
In order to unclutter the notation,
we shall sometimes omit the index A on the analytic-basis time variable,
hoping that this will not give rise to confusion. The difference between $t_A$
and $t$ should be
taken into account when rewriting the analytic superfields in the central basis.

Using the transformation laws \p{Vtrans} and \p{Stranharm} with keeping
in mind the
definitions \p{defLambda} it is straightforward to find that
$V^{++}(\zeta, u)$ transforms under $D(2,1;\alpha)$ like
\be
\delta' V^{++} = 2\Lambda\, V^{++}\,.\label{transfV++}
\ee
Using the properties \p{harmDtran}, \p{genLambd} it is also easy to check
the covariance of the harmonic constraint
(\ref{Hconstr}b) with respect to these transformations. For further use
we shall define non-analytic harmonic superfields $V^{--}$ and $V^{+-}$ by
\be
V^{+-} = \sfrac12 D^{--}V^{++}\,, \quad V^{--} = \sfrac12 (D^{--})^2 V^{++}\,
\ee
the transformation properties of which can be easily found using \p{transfV++}
and \p{harmDtran}:
\bea
&&\delta' V^{+-} = (2\Lambda -D^{--}\Lambda^{++})\,V^{+-}
+(D^{--}\Lambda^{++})V^{++}\;,  \nn
&&\delta' V^{--} = 2(\Lambda - D^{--}\Lambda^{++})\,V^{--} +[4D^{--}\Lambda
- (D^{--})^2\Lambda^{++}]\,V^{+-}\;.
\eea
It is worth noting that
\bea
&& V^{++}V^{--} - (V^{+-})^2 = \sfrac12 V^{ik}V_{ik}\,, \label{V2} \\
&& \delta' [V^{++}V^{--} - (V^{+-})^2] =
(2\Lambda - D^{--}\Lambda^{++})[V^{++}V^{--} - (V^{+-})^2]\;, \label{V2Htran}
\eea
which may be checked to coincide with the transformation law \p{Vtrans2}.

So much for the harmonic superspace formulation of the $({\bf 3,4,1})$ multiplet
$V^{++}$. In the next
section we shall demonstrate how the sigma-model actions \p{kin1}, \p{neq-1},
\p{eq-1} can also be rewritten
in harmonic superspace. We shall then construct the most general manifestly
$N{=}4$ invariant superpotential
term and its superconformal version as integrals over the analytic superspace
\p{analSS}.
In the remainder of the present section we briefly dwell on some direct
generalizations of $V^{++}$.

One may consider a general analytic superfield $q^{(+n)}$ with harmonic
U(1) charge equal to $+n$
and impose on it the same constraint as (\ref{Hconstr}b),
\be
D^{++} q^{(+n)} = 0\, \label{genHconstr} \;.
\ee
It is easy to check that, unless $n=0$, this constraint defines an off-shell
$N{=}4, d{=}1$ multiplet
with the field content $[{\bf2(n{+}1), 4n, 2(n{-}1)}]$ (for even $n$ one
may halve this content by imposing a reality condition).
For $n\le0$, eq. \p{genHconstr}
constrains $q^{(+n)}$ to be a constant or to vanish, in accord with
one of the basic principles of the harmonic superspace formalism \cite{book}:
$D^{++}f^{(-n)} = 0\; \Rightarrow \; f^{(-n)} = 0$. The superfields $q^{(+n)}$
for any $n>0$ are superconformal, in the sense that the whole $D(2,1;\alpha)$
at any $\alpha$ admits
a self-consistent realization on them. This realization uniquely follows from
requiring
\p{genHconstr} to be covariant with respect to the $D(2,1;\alpha)$
transformation \p{harmDtran}:
\be
\delta' q^{(+n)} = n\,\Lambda \,q^{(+n)}
\ee
(one should take into account here the property $\Lambda^{++} = D^{++}\Lambda$).
Note that
the first component of $\Lambda $ is the parameter of dilatations times $\alpha/2$,
so the dilatation weight of $q^{(+n)}$ (in the mass units) turns out to be
strictly related
with its harmonic U(1) charge:
\be
d(q^{(+n)}) = - \frac{\alpha}{2}\,n\,.
\ee
In particular, the dilatation weight of $V^{++}{\equiv}q^{(+2)}$ is $-\alpha$.

It is interesting to consider the particular value $n{=}1$. In this case we deal with the
superfield
$q^{+}(\zeta, u)$ having the off-shell field content $({\bf 4,4,0})$ ---
the only possibility for an off-shell representation of $N{=}4, d{=}1$ supersymmetry
without any auxiliary fields!
It is convenient to group $q^+$ and $\widetilde{q^+}$ into a doublet of an extra
(the so-called Pauli-G\"ursey) SU(2),
\cite{book})
\be
(q^+_a) = (q^+, -\widetilde{q}^+) \qquad\textrm{so that}\quad
\widetilde{q^+_a} = q^{+a} = \epsilon^{ab}q^+_b
\qquad\textrm{for}\quad a,b=1,2 \;. \label{qreality}
\ee
The explicit solution of \p{genHconstr} for this case, i.e. of
\be
D^{++} q^{+a} = 0\,, \label{qconstr}
\ee
reads
\be
q^{+a}(\zeta, u) = f^{ia}(t_A)u^+_i + \theta^+ \chi^a(t_A) +
\bar\theta^+ \bar\chi^a(t_A) +
2i \theta^+\bar\theta^+ \dot{f}^{ia}(t_A) u^-_i\,, \label{qsolv}
\ee
where
\be
\overline{(f^{ia})} = \epsilon_{ab}\epsilon_{ik} f^{kb}\,,
\quad \overline{(\chi^a)} = \bar\chi_a\,,
\ee
as a consequence of the reality property \p{qreality}. This $N{=}4, d{=}1$ multiplet
already appeared  in \cite{papa2,gps}
in component and in $N{=}1$ superfield form and also in \cite{hp,hull} as
a constrained $N{=}4$ superfield in the standard
$N{=}4, d{=}1$ superspace. Its relation with some other off-shell $N{=}4, d{=}1$ multiplets
also having 4
bosonic and 4 fermionic components, in particular with the above $({\bf 3,4,1})$ one,
was discussed
at the algebraic level in \cite{pt}. We see here that this field has a very simple
description as a constrained
superfield in analytic harmonic $N{=}4, d{=}1$ superspace. For later use
we spell out its
$D(2,1;\alpha)$ transformation law,
\be
\delta' q^{+a} = \Lambda\, q^{+a}\,.\label{qtran}
\ee
Also note that in the central basis eq. \p{qconstr} and the Grassmann analyticity
conditions imply the standard ${\bf R}^{1|4}$ constraints
\be
q^{+a}(z,u)  = q^{ia}(z)u^+_i\qquad\textrm{and}\qquad
D^{(i}q^{k)a} = \bar D^{(i}q^{k)a} = 0\,.\label{centrq}
\ee
They can easily be shown to be equivalent to those employed in \cite{hp,hull}.

Both by its field content and by its superconformal transformation properties
the superfield $q^{+a}$ resembles
the analytic hypermultiplet superfield of $N{=}2, d=4$ Poincar\'e supersymmetry.
The crucial difference between
the two is that in the hypermultiplet case the constraint \p{qconstr} puts this
superfield on shell,
while in our case it defines an off-shell multiplet
without any further dynamical constraints. Another difference is that the
transformation \p{qtran}
is a realization of the supergroup $D(2,1;\alpha)$ which cannot be obtained
as a reduction of the
$N{=}2, d=4$ superconformal group SU($2,2|2$). Both these crucial distinctions
are of course specific
features of $N{=}4, d{=}1$ supersymmetry.

In what follows we shall focus on the multiplets $V^{++}$ and $q^{+a}$, leaving
for the future the study of
possible implications of the supermultiplets $q^{(+n)}$ with $n>2$ in $N{=}4$ SQM models.
For completeness,
we mention that two further $N{=}4, d{=}1$ superfields have been used in constructing $N{=}4$
SQM models, those comprising
a $({\bf 1,4,3})$ multiplet \cite{ikl2,7,ikp} and a $({\bf 2,4,2})$ multiplet
\cite{fr,ikl2,ikp,bn,bgk}.
The first superfield is real and obeys constraints which are bilinear in spinor
derivatives and so cannot be interpreted as Grassmann analyticity conditions,
while the
second one is simply the chiral $N{=}4, d{=}1$ superfield. Clearly, the natural
superspaces to deal with
these multiplets are, respectively, the standard and chiral $N{=}4, d{=}1$ superspaces,
but not the harmonic superspace.

\setcounter{equation}{0}
\section{Harmonic superfield actions for $V^{++}$}
%and $q^{+a}$}
%\subsection{The $V^{++}$ actions}

\noindent{\it 5.1 Sigma-model type actions}
\vspace{0.2cm}

We begin by discussing the sigma-model actions for $V^{++}$.
When considering the most general sigma-model type action for $V^{++}$,
the harmonic superspace approach does not bring any new features. Indeed,
by dimensional arguments,
the corresponding action is naturally written as an integral over the
whole harmonic superspace,
\be
S_{(V)} = -\gamma \int du dt d^4\theta \,{\cal L}(V^{++}, V^{+-}, V^{--}, u)\,.
\label{HarmV1}
\ee
Then, choosing the central basis, substituting $V^{\pm\pm} = V^{ik}u^\pm_iu^\pm_k$
and
$V^{+-} = V^{ik}u^+_iu^-_k$ into \p{HarmV1} and integrating over harmonics
we return to the general action
\p{kin1} of the ${\bf R}^{1|4}$ superfield $V^{ik}(z)$ with
$$
L(V) = \int du {\cal L}(V^{++}, V^{+-}, V^{--}, u)
$$
(or its $3n$-dimensional target-space generalization in the case of
several $V^{ik}$ multiplets).

One may wonder whether self-consistent sigma-model type actions for $V^{++}$
can be written in
the harmonic analytic superspace ${\bf AR}^{1+2|2}$ \p{analSS}.
To single out such a subclass in
the set of general actions \p{kin1}, let us start with the free action corresponding
to $L(V) = 1/6\, V^{ik}V_{ik}$ and
rewrite it as the following integral over the whole harmonic
superspace ${\bf HR}^{1+2|4}$:
\be
S^{free}_{(V)} = -\sfrac16 \int dt d^4\theta \,V^2 =
-\sfrac14\int du dt_A d^4\theta\, V^{++}(D^{--})^2 V^{++}\,, \label{HarmV2}
\ee
where we made use of the relations \p{V2} and integrated by parts with respect
to harmonics.
Using the relations \p{measHA} and \p{Alg1}, we can further rewrite \p{HarmV2}
as an integral over the analytic
superspace,\footnote{
Hereafter we omit the index $A$ on the time derivative
because $\partial_{t_A} = \partial_t$.}
\be
S^{free}_{(V)} =
i\int du\,d\zeta^{--}\, V^{++}\left( D^{--}\partial_t +
\sfrac{i}{2}D^-\bar D^{-} \right)V^{++}\,.\label{HarmV3}
\ee
Using \p{Alg1}, it is easy to check that the differential operator in \p{HarmV3}
commutes with $D^{+}$ and $\bar D^{+}$ and so preserves the analyticity.

The evident analyticity-preserving nonlinear sigma-model extension of \p{HarmV3}
for $n$ superfields $V^{++}_B\,, \;B=1,2,\ldots, n\,,$ is as follows,
\be
S_{(V)}^{n\ \sigma} = i\int du\,d\zeta^{--}\, {\cal L}^{++\,B}(V^{++}_1,\ldots, u)
\left( D^{--}\partial_t + \sfrac{i}{2}D^-\bar D^{-} \right)V^{++}_B\;.\label{HarmV4}
\ee
The target geometry prepotential ${\cal L}^{++B}(V, u)$ is an arbitrary charge
2 function of $V^{++}_A$
and explicit harmonics. The bosonic sigma-model action is easily computed
to be of the general form
\p{bosN}, \p{defHG} where the tensors $H^{AB}$ and $G^{[A B]}_{(ik)}$ are
expressed as the following harmonic integrals,
\be
H^{AB} = \int du\; \partial_{++}^{(A}{\cal L}^{++\,B)}(v^{++}, u)\,,
\quad  G^{[A B]}_{(ik)} =
\int du\; \partial_{++}^{[A}{\cal L}^{++\,B]}(v^{++}, u)\,u^+_{(i}u^-_{k)}\,,
\label{spec}
\ee
where $\partial^A_{++} =\partial/\partial v^{++}_A$. The corresponding function
$L(v)$ is given by
\be
L(v) = \sfrac12 \int du\; {\cal L}^{++\,B}(V^{++}, u)V^{--}_B\,.\label{specL}
\ee
Due to the fact that all geometric objects in this special case are
expressed through the single
harmonic unconstrained prepotential ${\cal L}^{++ A}$, there appear some relations
between them
and also further restrictions.
In particular, it is easy to check that $H^{AB}$ and $G^{[AB]}_{(ik)}$ obey
the generalized harmonicity condition
\be
\Delta^{AB} H^{CD} = \Delta^{AB} G^{[C D]}_{(ik)} = 0\,.\label{constrAdd}
\ee
In the $3$-dimensional case, we are facing the conformally flat
3-dimensional metric \p{bos1},
with $H(v)$ being now some harmonic function,\footnote{
We hope that the use of the term `harmonic' in two different
meanings is not too confusing.}
\be
\Delta H(v) = 0\,.\label{constrA1}
\ee
Presently, we do not fully understand this special $3n$-geometry. Perhaps
it can be obtained as
a reduction of some strong $4n$-dimensional HKT geometry \cite{gps}. Recall
that it is specified by
the requirement of $N{=}4, d{=}1$ supersymmetry and, in addition, by the assertion
that the sigma model action
admits a representation in analytic harmonic $N{=}4, d{=}1$ superspace
$\{\zeta, u\}$, with the Lagrangian
being a local function of the analytic superfields $V^{++}_A$.\\

\noindent{\it 5.2 Analytic superpotentials}
\vspace{0.2cm}

The actual virtue of the harmonic superspace emerges
in the opportunity to write down the general manifestly $N{=}4$
supersymmetric superpotential term for $V^{++}$.
In the case of a single $V^{++}$ it is given by the following integral
over the analytic superspace,
\be
S_{(V)}^{sp} = -i\gamma{}'\int du d\zeta^{--} \,L^{++}(V^{++}, u) \;. \label{SpotV}
\ee
This superpotential is manifestly $N{=}4$ supersymmetric since the Lagrangian
is defined on the analytic
superspace which is closed under $N{=}4$ supersymmetry and the integral
is taken over this superspace.\footnote{This
should be contrasted with a non-manifestly supersymmetric way of writing
the superpotential as an integral
over the chiral $N{=}4, d{=}1$ superspace with a non-holomorphic Lagrangian
density \cite{hp}.} Thus $L^{++}$ can be
an arbitrary function of its arguments, and there is no need to care
about the conditions which the bosonic background
should satisfy for $N{=}4$ supersymmetry to be valid. Actually these
conditions can now be {\it derived\/} from
\p{SpotV}, passing there to components and considering the bosonic sector.
Substituting the analytic basis form
\p{solv(b)} for $V^{++}$ into \p{SpotV} and integrating over
$\theta^+, \bar\theta^+$, we find the general structure
of the component bosonic sector of \p{SpotV},
\be
\hat{S}^{sp}_{(V)} = \gamma{}' \int dt \left\{F\,{\cal V}(v) +
\dot{v}^{ik}\,{\cal A}_{ik}(v)\right\}\,, \label{half}
\ee
where the background scalar `half-potential' ${\cal V}$ and the magnetic
one-form potential ${\cal A}_{ik}$ are given
by the following harmonic integrals,
\be
{\cal V}(v) = \int du\, \frac{\partial L^{++}}{\partial v^{++}}\,, \qquad
{\cal A}_{ik}(v) = 2 \int du\, u^+_{(i}u^-_{k)}\,
\frac{\partial L^{++}}{\partial v^{++}}\,,\qquad
v^{++} = v^{ik}u^+_iu^+_k\,. \label{VA}
\ee
The term `half-potential' is related to the fact that the genuine scalar potential
$W$ appears
as the result of eliminating the auxiliary field $F(t)$ in the sum of the sigma-model
action \p{bos1}
and \p{half} as
\be
W(v) = -\sfrac12\frac{\gamma'{}^2}{\gamma} \frac{({\cal V}(v))^2}{H(v)}\;.
\label{poten}
\ee

We observe that both objects, ${\cal V}(v)$ and ${\cal A}_{ik}$, are expressed
through the same
analytic `prepotential' $L^{++}(v^{++}, u)$ which is required to have harmonic
U(1) charge $+2$
but is unconstrained otherwise. Choosing it at will,
we always obtain a `half-potential' and a background vector potential compatible
with $N{=}4, d{=}1$
supersymmetry. Inversely, the representation \p{VA} allows one to find the most
general constraints which
${\cal V}$ and ${\cal A}_{ik}$ should obey in order to admit an $N{=}4$ supersymmetric
extension. These
are easily computed to be
\bea
&&\partial_{ik}{\cal A}_{lt} - \partial_{lt}{\cal A}_{ik} =
\epsilon_{il}\,\partial_{kt}{\cal V} + \epsilon_{kt}\,\partial_{il}{\cal V}
\qquad\textrm{and}\qquad \Delta {\cal V} =0\,.
%\qquad \Delta = \frac{\partial^2}{\partial v^{ik}\partial v_{ik}}\;.
\label{monop}
\eea

This system is recognized as the static-solution (monopole) ansatz for a self-dual
Maxwell field in ${\bf R}^4$.
In fact, the set of equations \p{monop} is the same as in the famous Gibbons-Hawking
multi-center ansatz for 4-dimensional
hyper-K\"ahler metrics \cite{gh,egh}. Leaving aside the issue
of boundary condition in the target space, any solution of these equations,
up to gauge freedom,
give us a `half-potential' and a background vector potential admissible from
the point of view of $N{=}4$
supersymmetry. The simplest choice is the `Fayet-Iliopoulos term' \cite{bepa}
\be
L^{++} = c\,V^{++} \quad\textrm{with}\quad c=\textrm{const}\in{\bf R}
\qquad\Longrightarrow\qquad {\cal V} = c\,, \quad {\cal A}_{ik} = 0 \,,
\ee
and the only effect of adding such a superpotential term consists in
producing a scalar potential
\be
W \sim -\frac{c^2}{H(v)}
\ee
in the case of a non-trivial metric function $H(v)$.

A more complicated and interesting example is the well known multi-center solution
\be
{\cal V} = c_0 + \sum_A \frac{c_A}{|{\bf v} - {\bf v}_A|}\,,\label{multipot}
\ee
%{\bf [ v should be defined here ]} \newline
where ${\bf v} = (v^{ik})\,, {\bf v}\cdot{\bf v} = v^{ik}v_{ik}$. The corresponding
${\cal A}_{ik}$ can be straightforwardly found, but we do not give
them here.
%\footnote{The explicit form of the vector potentials corresponding to
%\p{multipot} can be found e.g. in \cite{}.}
The  notorious example of this kind is the spherically symmetric solution
\be
{\cal V} = c_0 + |{\bf v}|^{-1} \label{sphersymm}
\ee
for which
\be
F_{ik, lt} := \partial_{ik}{\cal A}_{lt} - \partial_{lt}{\cal A}_{ik} = -
(\,\epsilon_{il}v_{kt} + \epsilon_{kt}v_{il}\,)\,|{\bf v}|^{-3}\;.
\ee
This is the Dirac magnetic monopole. One can easily
find the corresponding prepotential $L^{++}(v^{++}, u)$ and compute the
relevant vector potential ${\cal A}_{ik}$ via eq. \p{VA}. We shall do this below,
after discussing $N{=}4$ superconformally invariant superpotentials.

Note that the $N{=}4$ supersymmetry constraints \p{VA} were derived for the first
time in \cite{cromrit} in an on-shell Hamiltonian approach (see also \cite{andr1}).\\

\noindent{\it 5.3 N=4 superconformally invariant superpotential}
\vspace{0.2cm}

In order to find
an $N{=}4$ superconformally invariant potential for $V^{++}$ we
shall follow a strategy which was applied for constructing the $N{=}2, d{=}4$ analytic
harmonic superfield action
of the improved tensor multiplet in \cite{gio1} and the $N{=}(4,4), d=2$ harmonic
superspace action of the $N{=}(4,4), SU(2)$
WZW model in \cite{IvSu}. Actually, the action we are going to construct
is the true $d{=}1$
analog of the general harmonic analytic action of the $N{=}2, d=4$ improved
tensor multiplet. The
difference lies, however, in the fact that the latter produces the sigma-model
type action for physical bosons,
with two derivatives on the latter, while in the case under consideration
we end up
with the sum of the scalar potential and the coupling to the background
vector potential, with only one
time derivative on the physical bosonic field $v^{ik}$.

The trick of \cite{gio1} adapted to the given case works as follows.
Let us split
\be
V^{++} = \hat{V}^{++} + c^{++}\,, \label{spl}
\ee
where
%\newline
%{\bf [ $c^{+-}$ and $c^{--}$ should be defined here ]}
\be
c^{\pm\pm}:= c^{ik}u^\pm_iu^\pm_k\,, \quad c^{+-}:= c^{ik}u^+_iu^-_k\,,\quad
c^{++}c^{--} - (c^{+-})^2 = \sfrac{1}{2}c^{ik}c_{ik} = 1\,, \label{defC}
\ee
and the constant 3-vector $c^{ik}$ satisfies the same reality condition as $V^{ik}$.
The $D(2,1;\alpha)$ transformations
\p{Stranharm}, \p{transfV++} imply for $\hat{V}^{++}$ the inhomogeneous
transformation law
\be
\delta' \hat{V}^{++} = 2\Lambda (\hat V^{++} + c^{++}) - 2\Lambda^{++} c^{+-}\,.
\label{transfhat}
\ee
Then, one searches for $L^{++}_{conf}(V^{++}, u)$ as a power series in
$X:=c^{--}\hat{V}^{++}$:
\be
L^{++}_{conf} = \hat{V}^{++}\sum_{n=0}^{\infty}b_n\,X^n
=: \sum_{n=0}^\infty L^{++}_{(n)}\,.\label{confprob}
\ee
The basic idea is to take advantage of the inhomogeneity of the transformation
law \p{transfhat} in order
to cancel the variations between the adjacent terms in the sum \p{confprob},
integrating by parts
with respect to $D^{++}$ and using the relations
\be
D^{++}\hat{V}^{++} = 0\,, \;\; D^{++}c^{++} = 0\;, \;\; D^{++}c^{--} =
2 c^{+-}\;,\;\; D^{++}c^{+-} = c^{++}\,.
\label{cConstr}
\ee
As the necessary condition for such cancellations, there will appear recurrence
relation involving the
coefficients in \p{confprob} which will be used to restore the precise
functional form of $L^{++}_{conf}$.

Keeping in mind that the integration measure of analytic superspace
is invariant under $D(2,1;\alpha)$ transformations,
the variation of the first term in \p{confprob}, up to a total harmonic derivative,
reads
\be
\delta' L^{++}_{(0)} = 2b_0\,\Lambda \hat{V}^{++}\;. \label{0var}
\ee
Analogously, the variation of the next term can be cast in the form
\be
\delta' L^{++}_{(1)} = 4b_1 \Lambda c^{++}c^{--} \hat{V}^{++}
                     + 4b_1 \Lambda c^{--} (\hat{V}^{++})^2 \;. \label{1var}
\ee
Using the relations \p{defC} and \p{cConstr}, it is easy to find
\be
c^{++}c^{--} =\sfrac{1}{12} (D^{++})^2(c^{--})^2 + \sfrac{2}{3}\,.
\ee
Substituting this into the first term of \p{1var}, we find that,
up to a total derivative, it reduces to
\be
\sfrac{8}{3} b_1 \Lambda\, \hat{V}^{++}\,.
\ee
Comparing this with \p{0var}, we find the condition of their cancellation to be
\be
b_1 = -\frac{3}{4} b_0\,.
\ee
Similarly, the remaining part of $\delta' L^{++}_{(1)}$ is cancelled by the
inhomogeneous part of
the variation $\delta' L^{++}_{(2)}$, provided that
\be
b_2 = -\frac{5}{6}\,b_1\,.
\ee
Continuing this procedure by induction and comparing the variations
$\delta' L^{++}_{(n-1)}$ and $\delta' L^{++}_{(n)}$,
we find the recurrence relation
\be
b_n = -\frac{2n+1}{2n +2}\,b_{n-1}\,.
\ee
Solving the latter, one obtains
\bea
L^{++}_{conf}
&=& b_0\,\hat{V}^{++}\sum_{n=0}^\infty \frac{(2n+1)!}{(n+1)!\;n!}
\left(-\sfrac{1}{4}X\right)^n \nn
&=& 2b_0\,\frac{\hat{V}^{++}}{\sqrt{1 + c^{--}\hat{V}^{++}}
\left(1 + \sqrt{1 + c^{--}\hat{V}^{++}}\right)}\,. \label{Lconf}
\eea
Now it is straightforward to explicitly check that the resulting action
\be
S^{sp\ conf}_{(V)} = -ib_0\,\int du d\zeta^{--}\,
\frac{2\,\hat{V}^{++}}{\sqrt{1 + c^{--}\hat{V}^{++}}
\left(1 + \sqrt{1 + c^{--}\hat{V}^{++}}\right)} \label{confSpot}
\ee
is invariant under \p{transfhat} up to a total harmonic derivative.
This can be done with making use
of the formula
\be
\delta' S^{sp\ conf}_{(V)} = -ib_0\,
\int du d\zeta^{--}\,\frac{\delta'\hat{V}^{++}}{
\left(\sqrt{1 + c^{--}\hat{V}^{++}}\right)^3}\,. \label{var}
\ee

Following \p{VA},
the half-potential and vector potential specifying the bosonic sector
of \p{confSpot} are given by the following
harmonic integrals,
\bea
&& {\cal V}^{conf} = \int du \frac{1}{\left(\sqrt{1 + c^{--}\hat{v}^{++}}\right)^3}\,,
\label{confpot} \\
&& {\cal A}_{ik}^{conf} = 2 \int du\;
\frac{u^+_{(i}u^-_{k)}}{\left(\sqrt{1 + c^{--}\hat{v}^{++}}\right)^3}
\label{Aconf}
\eea
($b_0$ was identified here with the overall normalization constant
$\gamma{}'$ and detached). The harmonic
integral \p{confpot} already appeared (in another context)
in \cite{gio1} and was computed
there. Using this result we find
\be
{\cal V}^{conf} = \frac{\sqrt{2}}{\sqrt{v^{ik}v_{ik}}} =
\frac{\sqrt{2}}{|{\bf v}|}\,.\label{qul}
\ee
We notice that the constant triplet $c^{ik}$ could be absorbed into
$v^{ik} = \hat{v}^{ik} + c^{ik}$
after the $u$-integration was performed, indicating that the $c^{ik}$
are moduli of the theory.
Yet, they still appear explicitly in ${\cal A}_{ik}^{conf}$. The latter
can be calculated by
choosing  in \p{Aconf} an explicit parametrization of harmonics, e.g.
by Euler angles, and performing
the integral over $S^2$. However, it is simpler to directly restore
${\cal A}_{ik}^{conf}$ from
the general constraint \p{VA}:
\be
{\cal A}_{ik}^{conf} = -\sqrt{2}\frac{c_i^pv_{pk} +
c_k^pv_{pi}}{[({\bf v}\cdot {\bf c}) + \sqrt{2}|{\bf v}|]|{\bf v}|}\,.
\label{confVect}
\ee
This is the potential of a Dirac magnetic monopole, with $c^{ik}$ parametrizing
the singular Dirac string.
Thus, in the manifestly $N{=}4$ supersymmetric formulation of the conformally
invariant superpotential these parameters
arise already at the superfield Lagrangian level. However, the complete
action does not depend on them,
because of its scale invariance and its invariance under the `conformal' SU(2)
(the one which acts on the indices $i,j$).
The same is valid for the bosonic part of the action: when \p{confVect}
is substituted
into \p{half}, all terms with manifest $c$-dependence are reduced
to a total $t$-derivative \cite{ikl}. The bosonic
sector of $L^{++}_{conf}$ precisely coincides with that derived in \cite{ikl}
from the formulation
of $N{=}4$ superconformal mechanics in the standard $N{=}4, d{=}1$
superspace and in terms of $N{=}2$ superfields.

It is worth pointing out once more that the possibility to write the
superpotential term in
manifestly $N{=}4$ supersymmetric form (both for the generic prepotential $L^{++}$
and
for the superconformally invariant one $L^{++}_{conf}$) is offered only
by the analytic harmonic
$N{=}4, d{=}1$ superspace. One can of course rewrite them as integrals over
the complete $N{=}4$
superspace, either in terms of the prepotential for $V^{ik}$ or
with explicit $\theta$ s,
like this has been given for the superconformal case in \cite{ikl}.
However, such a
representation, as opposed to the formulation in ${\bf AR}^{1+2|2}$,
lacks manifest supersymmetry (or makes one to care about some superfluous gauge
invariances)
and does not suggest any hint how to generalize the superconformally invariant
superpotential
(or the FI term) to the generic case. \\

\noindent{\it 5.4 Further examples of superpotentials}
\vspace{0.2cm}

Note that the general spherically symmetric half-potential \p{sphersymm}
also yields the expression \p{confVect} for ${\cal A}_{ik}^{conf}$, but for
any non-zero constant $c_0$ it breaks conformal
invariance, and the corresponding $L^{++}$ does not respect $N{=}4$ superconformal symmetry.
Such a superpotential is
obtained by adding to $L^{++}_{conf}$ the non-conformal piece $c_0 V^{++}$, i.e.
\be
L^{++}_{TN} = c_0\,V^{++} + \sfrac{1}{\sqrt{2}}\,L^{++}_{conf}\,. \label{TN}
\ee
Though $L^{++}_{TN}$ breaks the whole $N{=}4$ superconformal group, it is still invariant
under the SU(2) subgroup of the
latter which affects the harmonics and the doublet indices of the original
$\theta_i,\bar\theta^k$ and acts as rotations on $v^{ik}$. It corresponds to the choice
\be
\Lambda_{su(2)} = \lambda^{(ik)}u^+_iu^-_k = \sfrac12 D^{++}\lambda^{(ik)}u^-_iu^-_k
\label{LSU(2)}
\ee
in the transformation laws \p{Stranharm}, \p{transfhat}. Varying the first term in \p{TN} as
$\delta_{su(2)}V^{++} = 2\Lambda_{su(2)}V^{++}$, using \p{LSU(2)} and integrating by parts
with respect
to $D^{++}$, one proves the SU(2) invariance of this term. The second term is invariant under
the whole $D(2,1;\alpha)$
and hence under any of its subgroups.

The superpotential \p{TN} is a $d{=}1$ analog of one of the two forms of the Taub-NUT
sigma-model Lagrangian
in analytic harmonic $N{=}2, d=4$ superspace \cite{book}: both are specified by the Taub-NUT
potential \p{sphersymm}.\footnote{
To avoid a possible misunderstanding, let us point out that in the former case one deals
with a 4-dimensional
hyper-K\"ahler manifold possessing SU(2)$\times$ U(1) isometry group. In the multi-center
ansatz the U(1) isometry
is realized as a shift of 4th coordinate and affects neither the Taub-NUT potential
nor the relevant vector
potential which are defined on ${\bf R}^3$.  In our case we are dealing at once with
a $3$-dimensional bosonic manifold
and the Taub-NUT scalar and vector potentials defined on it; so the maximal invariance
group of \p{TN} is SU(2).}
It is interesting to find the superpotential producing the well known two-center potential
for the Eguchi-Hanson metric
as the associated half-potential ${\cal V}$. Such an $L^{++}_{EH}$ is given by
\bea
L^{++}_{EH} &=& \frac{c_1\,V^{++}}{\left( 1 + \sqrt{1 - V^{++}a^{--}}\right)
\sqrt{1 - V^{++}a^{--}}} \nn
&+& \frac{(V^{++} -c^{++})}{\left( 1 + \sqrt{1 + (V^{++}-c^{++})c^{--}}\right)
\sqrt{1 + (V^{++} -c^{++})c^{--}}}\,, \label{EH1}
\eea
where $a^{--} = a^{ik}u^-_iu^-_k\,, \;\;a^{ik}a_{ik} = 2$, and it leads to
\be
{\cal V}_{EH} = \int du \,\frac{1}{(\sqrt{1 + (v^{++} -c^{++})c^{--}})^3} + c_1
\int du \,\frac{1}{(\sqrt{1 - v^{++}a^{--}})^3}\,. \label{EH2}
\ee
These harmonic integrals are computed like in the previous case, yielding,
up to an overall factor,
\be
{\cal V}_{EH} = \frac{1}{|{\bf v}|} + c_1 \frac{1}{|{\bf v} - {\bf a}|}\,. \label{EH3}
\ee
This half-potential has poles located at ${\bf v} = 0\,, \;{\bf v}={\bf a}$ and, up
to normalization and shifts of $v^{ik}$, coincides with the standard potential
for the Eguchi-Hanson metric
\cite{egh}. Adding to \p{EH1} a term $\sim V^{++}$ adds a constant to \p{EH3}
and so produces a potential
for what is called the `double Taub-NUT metric' \cite{egh}.\\

\noindent{\it 5.5 Superpotentials in the 3n-dimensional case}
\vspace{0.2cm}

It is straightforward to generalize \p{SpotV} and \p{half} to the
case with $n$ superfields $V^{++}_A\,, \; A= 1,2,\ldots, n$, via
\be
S_{(V)}^{n\ sp} = - i\gamma' \,\int dud\zeta^{--} L^{++}(V^{++}_A, u)\,.
\label{SpotV2}
\ee
Its bosonic part reads
\be
\hat{S}^{n\ sp}_{(V)} = \gamma' \,\int dt \left\{F_B\,{\cal V}^B(v) +
\dot{v}_B^{ik}\,{\cal A}^B_{ik}(v)\right\}\,, \label{halfn}
\ee
with
\be
{\cal V}^B(v) = \int du\, \frac{\partial L^{++}}{\partial v^{++}_B}\,, \quad
{\cal A}^B_{ik}(v) = 2 \int du\, u^+_{(i}u^-_{k)}\,
\frac{\partial L^{++}}{\partial v^{++}_B}\;.\label{VAn}
\ee
The full potential can be read off by integrating out the auxiliary fields $F_B$
in the sum of the actions \p{bosN} and \p{halfn}.
Note that in the general case this procedure modifies as well the coupling
to the external gauge potential because
of a mixed term $\sim F_A \dot{v}^{ik}_B$ in \p{bosN}.

The constraints generalizing \p{VA} again follow from the explicit
expressions \p{VAn},
\bea
&&\partial^A_{ik}{\cal A}^B_{lt} - \partial^B_{lt}{\cal A}^A_{ik} =
\epsilon_{il}\,\partial^A_{kt}{\cal V}^B +
\epsilon_{kt}\,\partial^A_{il}{\cal V}^B\,, \nn
&& \Delta^{CD} {\cal V}^A =0\,, \quad \partial_{ik}^{[A}{\cal V}^{B]} = 0\,.
\label{constrgener}
%\qquad \Delta = \frac{\partial^2}{\partial v^{ik}\partial v_{ik}}\;. \label{II}
\eea
These constraints resemble those arising in the ansatz for a toric $4n$-dimensional
hyper-K\"ahler metric \cite{hktoric}
which generalizes the Gibbons-Hawking 4-dimensional one. The difference lies however
in the fact that
our half-potential ${\cal V}^A$ and one-form ${\cal A}_{ik}^A$ are vectors with respect
to the extra index $A$, while
in the ansatz of \cite{hktoric} the analogous objects are rank 2 symmetric tensors.

Finally, we would like to point out that the properties of
the sigma-model and superpotential parts of the full action for $V^{ik}$
do not correlate: in general, their only
common property is $N{=}4$ supersymmetry. For instance, one can choose the scalar
half-potential and one-form potential
to be spherically symmetric as in the examples above and, at the same time,
not assume such a symmetry for the
metric functions in \p{bos1}, \p{bosN}, or vice versa.

\setcounter{equation}{0}
\section{The actions for ${\bf q}^{{\bf+}}$}

\noindent{\it 6.1 General $q^+$ sigma-model action}
\vspace{0.2cm}

The constrained analytic $N{=}4$ superfield $q^{+a}$ is defined by eqs.
\p{qreality}-\p{centrq}.
The general sigma-model type off-shell action for $q^{+ a}$, like in
the case of $V^{++}$,
can be written either as an integral over ${\bf R}^{1|4}$ of a function
of the ordinary constrained superfields
$q^{ia}(z)$ (see \p{centrq}), or, equivalently, as an integral over
${\bf HR}^{1+2|4}$
of a function of the $q^{+a}$ with $q^{-a} = D^{--}q^{+a}$:
\be
S_{(q)} = -\int dtd^4\theta \, L(q) = -\int du\,dtd^4\theta \,L'(q^+, q^-, u)\,, \;
L(q) = \int du\, L'(q^+, q^-, u)\,.
\label{qgener1}
\ee
The simplest way to compute the bosonic sigma-model action is to find the
bosonic part of $q^{ia}$ (with fermions omitted)
and substitute it into \p{qgener1}. Passing to the central basis in the explicit
expression \p{qsolv} for $q^{+ a} = q^{ia}(z)u^+_i$
and suppressing the fermions, we find
\be
q^{ia}(z) = f^{ia}(t) - i \dot{f}{}^{ka}(t)(\theta_k\bar\theta^i
- \bar\theta_k\theta^i) +
\frac14\ddot{f}{}^{ia}(t)\theta^2\bar\theta^2\,. \label{shortq}
\ee
Then, for the bosonic core of \p{qgener1} we obtain the simple expression
\be
\hat{S}_{(q)} = \int dt\, \Delta L\, (\dot{f}\cdot\dot{f})\,,  \label{bosqgen}
\ee
where now $\Delta = \partial^2/ \partial f\cdot \partial f$ and ``$\cdot $''
denotes
contraction over the ${\bf R}^4$ indices, $x\cdot y \equiv x^{ia}y_{ia}$.
So we encounter a conformally-flat
geometry in ${\bf R}^4$. An extension to the case of several superfields
$q^{+ a}_A$, $A=1,\ldots, n$, is straightforward. One changes
$L(q) \rightarrow L(q_1,q_2,\ldots)$ and
obtains the bosonic action in the form
\bea
\hat{S}^{\;n}_{(q)} = -2\int dt \, g^{AB}_{a\,b}\,
(\dot{f}^{ja}_A\dot{f}^b_{j\,B})\,, \quad
g^{AB}_{a\,b} = \epsilon^{tj}
\frac{\partial^2\, L}{\partial f^{ta}_A\,\partial f^{jb}_B} \label{Gqgen}
\eea
The general characterization of this geometry was given in \cite{stro1,hp,gps,hull}.
As found in \cite{stro1,hull} (see also \cite{gp}), this geometry is in general
weak HKT and for special choices of the potentials $L(f)$ can be strong
HKT (torsion appears in the fermionic terms).

Like in the case of $V^{++}$, one can inquire whether for some special
$L(q)$ the above actions
admit a representation as integrals over the analytic harmonic superspace
$\{\zeta, u \}$ with
manifestly analytic superfield Lagrangians. Let us start from the free
theory which corresponds
to the choice $L(q) = \frac18 q \cdot q$
\be
S_{(q)}^{free} = -\sfrac18 \int dtd^4\theta\, (q \cdot q)
= -\sfrac14 \int du \,dtd^4\theta\, q^{+a}D^{--}q^+_a\,.
\label{free}
\ee
Passing, in the second form of this action, to the integral over the analytic superspace
with the help of \p{measHA} and \p{Alg1}, we find a very simple analytic superspace form of
this kinetic term,
\be
S^{free}_{(q)} =  \sfrac{i}{2} \int du d\zeta^{--}\,(q^{+a}\dot{q}{}^+_a)
\qquad \Longrightarrow \qquad
\hat{S}^{free}_{(q)} = \int dt\, (\dot{f}^{ia}\dot{f}_{ia})\,.\label{analitFree}
\ee
One can immediately generalize it to the interaction case. For  several $q^{+a}_A$,
this generalization reads
\be
S^{n\ \sigma}_{(q)} = i\gamma\, \int du d\zeta^{--}\,
{\cal L}^{+a A}(q^+, u)\,\dot{q}{}^+_{a A}\,,
\label{analitint}
\ee
where ${\cal L}^{+a A}(q^+, u)$ is some unconstrained analytic prepotential,
depending on $q^{+ a}_A$
and explicit harmonics. It is very easy to obtain the bosonic action,
\be
\hat{S}^{n\ \sigma}_{(q)} = -2 \gamma\, \int dt \left(\int du
\frac{\partial {\cal L}^{+A}_a}{\partial f^{+b}_B}
\right)\dot{f}^{ka}_A
\dot{f}^{b}_{k\,B}\,.\label{qspec}
\ee
Comparing with \p{Gqgen} we find (up to the normalization $\gamma$)
\be
\tilde{g}^{AB}_{a\,b} = \int du\, \frac{\partial {\cal L}^{+A}_a}{\partial f^{+b}_B}\,,
\quad \tilde{L}(f) =
\int du\, {\cal L}^{+a A}q^-_{a\,A}\,. \label{specbos}
\ee
where the tilde indicates that we deal with a particular case of \p{Gqgen}.
%{\bf [ why tildes here? ]} \newline
It immediately follows from \p{specbos} that such $\tilde{g}^{AB}_{a\,b}$ --
as a main specific feature -- satisfy the generalized harmonicity condition
\be
\Delta^{CD} \tilde{g}^{AB}_{a\,b} = 0 \qquad\textrm{with}\quad
\Delta^{CD} = \frac{\partial^2}{\partial_C\cdot \partial_{D}}\,.
\ee

In the 4-dimensional case ($n{=}1$) only the piece $\sim \epsilon_{ab}$ in
$\tilde{g}_{a\,b}$ contributes
to \p{qspec}, and the latter takes the conformally-flat form
\be
\hat{S}^{1\ \sigma}_{(q)} = -\gamma\, \int dt \,H(f)\,
(\dot{f}\cdot \dot{f})\,, \quad H(f) = \int du
\frac{\partial {\cal L}^{+a}}{\partial f^{+a}}\,, \quad
\Delta H(f) = 0\,. \label{qspec1}
\ee
This metric belongs to the class exhibited in \cite{chs,gps}, and it defines
a strong HKT geometry
associated with flat ${\bf R}^4$ as the relevant hyper-K\"ahler manifold
(in order to see this one should examine the torsion which appears in the
fermionic terms of \p{analitint}).

Thus, similarly to the $V^{++}$ case, we observe an interesting phenomenon.
Requiring off-shell $N{=}4$
supersymmetric sigma-model actions for $q^{+a}$ to be representable
in the analytic harmonic superspace,
with the Lagrangians being local functions of analytic $q^{+a}$ superfields,
imposes further restrictions
on the target-space geometry of such $d{=}1$ sigma models, in addition to those
which are already implied by $N{=}4$
supersymmetry and are automatically satisfied in the superfield approach. \\

\noindent{\it 6.2 N=4 supersymmetric off-shell coupling to the background
one-form}
\vspace{0.2cm}

One can easily write the analog of the superpotential \p{SpotV} for the
field $q^{+a}$,
\be
S_{(q)}^{sp} = -\sfrac{i}{2}\,\int dud\zeta^{--} \,{\cal L}^{++}(q^+, u)\,.
\label{Supq}
\ee
However, since the $({\bf 4, 4, 0})$ multiplet does not involve any auxiliary field,
\p{Supq}
does not give rise to any scalar potential and only provides a coupling to the external
background one-form.
It is the most general manifestly $N{=}4$ supersymmetric invariant of the appropriate dimension.
One can expect that the linearized version of the constraints on the background one-form
found in \cite{hp} by {\it requiring}
$N{=}4$ supersymmetry  automatically emerges from \p{Supq}.
\footnote{The authors of \cite{hp} considered some nonlinear
version of the $({\bf 4, 4, 0})$ multiplet constraints. For the moment we do not know how
to describe such a
nonlinear $({\bf 4,4,0})$ multiplet in harmonic superspace.}

For the simplest case of a single flied $q^{+a}$ one gets the following bosonic action,
\be
\hat{S}_{(q)}^{sp} = \int dt\,\dot{f}{}^{ia}\,{\cal A}_{ia}(f)\,,\quad {\cal A}_{ia}(f) =
\int du\,u^-_i \frac{\partial {\cal L}^{++}}{\partial f^{+a}}\,. \label{bosqA}
\ee
{}From the explicit representation for ${\cal A}_{ia}$ it is easy to check that
\be
{\cal F}_{kb\,ia} := \partial_{kb} {\cal A}_{ia} - \partial_{ia} {\cal A}_{kb} =
\epsilon_{ki}\,\int du \,
\frac{\partial^2 {\cal L}^{++}}{\partial f^{+a}\,\partial f^{+b}}
=: \epsilon_{ki} {\cal B}_{(ab)}\,, \label{selfd1}
\ee
which is a self-duality condition for the ${\bf R}^4$ abelian gauge field. The same
condition
has been obtained in \cite{hp}. In addition, \p{bosqA} implies that
\be
\partial^{ia}{\cal A}_{ia} = 0\,,\label{gauge1}
\ee
which can be interpreted as a choice of transversal gauge for ${\cal A}_{ia}$.
Together with the Bianchi identity
\be
\partial_{i}^{\;\;b}{\cal B}_{(ba)} = 0 \label{Bian}
\ee
the self-duality condition \p{selfd1} implies, as usual, that
\be
\partial^{ia}{\cal F}_{ia \,kb} = 0
\ee
which, with taking account of \p{gauge1}, yields
\be
\Delta {\cal A}_{ik} = 0\,.
\ee
This can be checked directly using \p{bosqA}. The constraints \p{selfd1}
and \p{gauge1} are the linearized
(adapted to the linear $({\bf 4,4,0})$ multiplet) form of the constraints
derived in \cite{hp}.

It is straightforward to generalize this consideration to several superfields
$q^{+a}_C$.
One should then allow for extra $q^{+a}_B$ superfields in ${\cal L}^{++}$ in \p{Supq}.
The relevant one-form
potential is given by the harmonic integral
\be
{\cal A}_{ia}^C = \int du \,u^-_i \frac{\partial {\cal L}^{++}}{\partial f^{+a}_C}\,.
\ee
It satisfies the following analogs of constraints \p{selfd1}, \p{gauge1}:
\bea
&&{\cal F}^{CD}_{ia\,kb} := \partial_{kb}^C {\cal A}_{ia}^D -
\partial_{ia}^D {\cal A}_{kb}^C = \epsilon_{ki}\,\int du \,
\frac{\partial^2 {\cal L}^{++}}{\partial f^{+a}_D\,\partial f^{+b}_C}
=: \epsilon_{ki} {\cal B}_{ab}^{CD}\,, \label{seldn} \\
&& \partial^{(B}\cdot {\cal A}^{C)} = 0\,,\label{gaugen}
\eea
as well as the generalized harmonicity condition
\be
\Delta^{AB}{\cal A}^C_{ia} = 0\,. \label{harmonn}
\ee

The last remark concerns possible scalar potentials of the fields $f^{ia}$.
As already mentioned, in the
pure $q^{+a}$ models such terms cannot appear because of absence of auxiliary
fields in the $({\bf 4, 4, 0})$ multiplet.
However, they can appear in mixed systems, e.g. with both $q^{+a}$ and
$V^{++}$ superfields involved. The simplest
possibility of this sort is to consider the manifestly $N{=}4$
supersymmetric analytic superpotential
\be
S_{(Vq)} = - i\int dud\zeta^{--}\,V^{++}\, L(q^+, u)\,, \label{mixedV}
\ee
which in the bosonic sector yields the term
\be
\hat{S}_{(Vq)} = \int dt\;F(t)\,\Bigl(\int du\; L(f^+, u)\,\Bigr)\quad
+ \quad \mbox{coupling to a one-form potential}\,.\label{halfpotq}
\ee
The half-potential ${\cal V}^{(q)}(f) = \int du L(f^{+a}, u)$ satisfies the four-dimensional
Laplace equation.
After $F(t)$ is integrated out from the sum of \p{halfpotq} and the free $V^{ik}$
action \p{HarmV3}, one ends
up with a scalar potential which depends only on $f^{ia}$. If instead of
the free $V^{ik}$ action we take the general
sigma-model type action \p{kin1}, the eventual scalar potential
will be a function of both $v^{ik}$ and $f^{ja}$.

One further possibility is to include into the game the odd-Grassmann
parity version of the $({\bf 4,4,0})$ multiplet.
It is described off shell by a fermionic analytic superfield
$\Psi^{+a}(\zeta, u)$ satisfying
the harmonic constraint
\be
D^{++}\Psi^{+a} = 0\,.
\ee
Therefore, it has the same $\theta^+, \bar\theta^+$ expansion
as $q^{+a}$ \p{qsolv}, namely
\be
\Psi^{+a}(\zeta, u) = \psi^{ia}u^{+}_i + \theta^+ \xi^a + \bar\theta^+ \bar\xi^a
+ 2i \,\theta^+\bar\theta^+ \dot\psi^{ia}u^-_i\,,
\ee
with the only (but essential) difference that $\psi^{ia}$ are now physical
fermions while $\xi^a, \bar\xi^a$ form
a complex doublet of bosonic auxiliary fields. So it can be called
the ${\bf 0,4,4})$ multiplet.
The free action of it has the following nice form,
\be
S_{(\psi)}^{free} = \sfrac{1}{2} \int du d\zeta^{--}\;\Psi^{+a}\Psi^+_a =
\int dt \left(i \psi^{ia}\dot{\psi}_{ia} +
\xi^a\bar \xi_a \right)\,.
\ee
When appropriately coupled to $q^{+a}$, this multiplet is also presumably capable
to produce scalar potentials for bosons
$f^{ia}$ upon eliminating the auxiliary fields $\xi^a, \bar\xi^a$.\\

\noindent{\it 6.3 Superconformally invariant $q^+$ actions}
\vspace{0.2cm}

It is natural to start the investigation of $D(2,1;\alpha)$ invariant actions
for $q^{+a}$
with the free action \p{free}. As before we shall be interested in invariance
under the transformations
of conformal supersymmetry. Taking into account the $D(2,1;\alpha)$
transformation law \p{qtran} and
the fact that $\partial/\partial t_A$ is properly transformed through partial
derivatives in the
analytic Grassmann coordinates, the superconformal variation of \p{free}
is as follows (up to a total
$D^{++}$ derivative),
\be
\delta' S^{free}_{(q)} =
-2(1{-}\alpha)\int du d\zeta^{--}\,(\epsilon^-\bar\theta^+
- \bar\epsilon^-\theta^+)\,q^{+a}\dot{q}{}^+_a\,.
\ee
We see that the free action is only invariant provided that $\alpha = 1$.

Nevertheless, one can define nonlinear sigma-model type $D(2,1;\alpha)$ invariant
actions for
$q^{+a}$ for any value of $\alpha $. Like in the case of $V^{++}$, such actions
admit a field theoretic
interpretation (i.e. contain a kinetic part) only under the assumption that
the bosonic fields
$f^{ia}$ start with some constant, $f^{ia} = \epsilon^{ia} + \ldots $. This means
that
$N{=}4$ superconformal symmetry is spontaneously broken and $q^{+a}$ is the corresponding
Goldstone superfield, analogous to the interpretation of $V^{++}$ in \cite{ikl}.
This issue is
considered in \cite{ikl2}. Here we wish to show that such superconformal actions
can be
constructed without any reference to the nonlinear realization formalism
used in \cite{ikl,ikl2}.

The idea behind the whole construction is very simple. Let us define a composite
$N{=}4$ analytic superfield
\be
{\cal V}^{++} = q^{+ a}a_{ab}q^{+b}\,\;\label{composite}
\ee
where $a_{ab}$ is a constant symmetric tensor which breaks the extra
Pauli-G\"ursey SU(2) (realized
on the indices $a, b$) down to some U(1) subgroup. We choose
\be
a^2 = a^{ab}a_{ab} = 1\,.
\ee
The superfield \p{composite} possesses all the properties of $V^{++}$.
Indeed, it satisfies
\be
D^{++}{\cal V}^{++} = 0\,,
\ee
as a consequence of the $q^{+a}$ defining constraint \p{qconstr}, and it
transforms under $D(2,1;\alpha)$ as
\be
\delta' {\cal V}^{++} = 2\lambda {\cal V}^{++}
\ee
as a result of the $q^{+a}$ transformation law \p{qtran}. Hence,
any action of $V^{++}$, including the
superconformally invariant ones \p{neq-1}, \p{eq-1}, upon substitution
$V^{++} \rightarrow {\cal V}^{++}$
will produce an $N{=}4$ supersymmetric action for $q^{+a}$.
This substitution applied to \p{neq-1}, \p{eq-1} evidently
produces the desired $D(2,1;\alpha)$ invariant actions for $q^{+a}$.
Using the central basis form of
\p{composite},
\be
{\cal V}^{ik}(z) = q^{ia}(z)a_{ab}q^{kb}(z)\,, \label{compos2}
\ee
and the relation
\be
{\cal V}^2 = \frac{1}{4}(q^2)(q^2)\qquad\textrm{where}\quad q^2 \equiv q^{ia}q_{ia}\,,
\ee
it is easy to recover the superconformally invariant actions of $q^{+a}$ as
\be \label{supconfq}
S^{conf}_{(q)} (\alpha) =\ \begin{cases}
\quad-\gamma\int dtd^4\theta\;(q^2)^{\frac{1}{\alpha}}
\qquad&\textrm{for}\quad\alpha\neq-1\\[8pt]
\quad\gamma'\int dtd^4\theta\;(q^2)^{-1}\ln (q^2)\qquad&\textrm{for}\quad\alpha
= -1\end{cases} \quad.
\ee
At $\alpha =1$ the upper version becomes free and we reproduce the previous result.
On the other hand,
the free action of $V^{ik}$ corresponding to $\alpha = \frac{1}{2}$ in \p{neq-1}
produces a nonlinear sigma-model
type action for $q^{ia}$ upon the substitution \p{compos2}.

The component bosonic action for \p{supconfq} is easy to compute
by the general formula \p{bosqgen}:
\be
\hat{S}_{(q)}^{conf} (\alpha) =\ \begin{cases}
\quad \gamma \sfrac{1+\alpha}{\alpha^2}\int dt \,
(f^2)^{\frac{1-\alpha}{\alpha}}\,(\dot{f}\cdot \dot{f})
\qquad&\textrm{for}\quad\alpha\neq-1\\[8pt]
\quad \gamma' \int dt \, (f^2)^{-2}\,(\dot{f}\cdot \dot{f}) \qquad&\textrm{for}\quad\alpha
= -1\end{cases}\quad. \label{bosconfq}
\ee
Except for the free $\alpha = 1$ case,
they make sense only under the assumption that the `vacuum value'
of the radial part of $q^{ia}$ is non-vanishing, i.e. $\langle q^2\rangle\neq 0$.
This means that the deviation
$\tilde{q}^2 = q^2 - \langle q^2\rangle$ is a dilaton with an inhomogeneous transformation law.
Thus, dilatation invariance
is spontaneously broken. Analogously, one
can see that the whole R-symmetry SU(2)$_R$ acting on the indices $i, j$ and rotating
the harmonics is
spontaneously broken, with the angular part of $q^{ia}$ being the corresponding
Goldstone fields.
This is to be contrasted with the field $V^{ik}$ which also involves
the dilaton as its radial part, but
breaks SU(2)$_R$ only down to U(1) \cite{ikl}. In both cases, another SU(2)
R-symmetry present in $D(2,1;\alpha)$
is linearly realized on the physical fermions and so is unbroken.
The superfield $q^{ia}$, similarly to $V^{ik}$,
can be derived from the appropriate nonlinear realization of $D(2,1;\alpha)$
as a Goldstone superfield \cite{ikol2}.
It is interesting that the Pauli-G\"ursey SU(2) group acting on the index
$a$ of $q^{ia}$ and explicitly
broken by the ansatz \p{composite} is actually restored in the
superconformal action \p{supconfq}.
The latter is thus manifestly SO(4) invariant.

For later use, it is of interest to recognize how the component fields
of ${\cal V}^{++}$
are expressed in terms of those of $q^{+a}$. Substituting the component
expressions \p{solv(b)}, \p{qsolv} into \p{composite}, we obtain
\be
{\cal V}_0^{ik} = f^{ia}a_{ab}f^{kb}\,, \; {\cal F} = 2\,(\,\dot{f}^{\;a}_i a_{ab} f^{ib}
+ i\chi^a a_{ab}\bar \chi^b\,)\,, \;
\psi^i = 2 \chi^aa_{ab}f^{ib}\,, \; \bar \psi^i =
2 \bar\chi^a a_{ab} f^{ib}\,.\label{compcompos}
\ee
%{\bf [ the last two are inconsistent] }\newline
The property that the $({\bf 4, 4, 0})$ multiplet can be obtained from
the $({\bf 3, 4, 1})$ one by replacing the auxiliary
field in the latter by a time derivative of some new scalar $d{=}1$ field
was mentioned in \cite{pt} as a
particular case of a more general phenomenon. The relations \p{composite},
\p{compcompos} can be
regarded as a nonlinear version of this correspondence.

We now can study what kind of the superpotential-type invariant for $q^{+a}$ the
substitution
\p{composite} produces from the superconformal superpotential \p{confSpot}.
The direct insertion of
the expressions \p{compcompos} into \p{half} with ${\cal V}_{conf}$
and ${\cal A}^{conf}_{ik}$
given by \p{confpot} and \p{Aconf} yields, up to a normalization factor,
\be
S_{(q)}^{sp\ conf}\;\;\Longrightarrow \;\;
\int dt \dot{f}^{ia}{\cal A}_{ia}^{conf}(f)\,, \label{confAf}
\ee
where
\be
{\cal A}^{conf}_{ia}(f) =
\frac{1}{\left[f^2 + 2 (f\cdot a\cdot f\cdot c)\right]}\left(f^k_ac_{ki}
- f^b_ia_{ab} \right)
\label{potq}
\ee
and
$$
f\cdot a\cdot f\cdot c = f^{ia}f^{kb}c_{ik}a_{ab}\,.
$$
However, calculating the curl of this vector potential, one finds the
disappointing result
\be
\partial_{kb}{\cal A}^{conf}_{ia} - \partial_{ia}{\cal A}^{conf}_{kb} = 0\,
\ee
i.e. ${\cal A}^{conf}_{ia}$ is pure gauge and \p{confAf} is a total derivative.
Note that the crucial
role in achieving this negative result is played by the term with the auxiliary
field in \p{half}: now it is proportional
to $\dot f^{ia}$, and its contribution combines with that from the magnetic
coupling in \p{half} to produce
the total derivative. An inspection of the fermionic terms in \p{confSpot}
with a composite $V^{++}$ also
shows that they vanish: the term $\sim \psi^i \bar\psi^k$ after
the substitution \p{compcompos} is cancelled out
by a similar term coming from the `auxiliary field' ${\cal F}$.

Thus, rather surprisingly, an $N{=}4$ superconformally invariant coupling
of $q^{ia}$ to the one-form target-space
potential does not exist. For the time being we do not fully understand
what stands behind this property.
Presumably, it is related to the absence of WZW terms for
the fully non-linearly realized
SU(2) group for which the angular part of the ${\bf R}^4$ vector $f^{ia}$
provides
a parametrization \cite{ikl2}. In the ${\bf R}^3$ case one deals with
the coset SU(2)/U(1), and
the $d{=}1$ WZW term associated with this U(1) is just the conformally invariant
coupling of $v^{ik}$ to the magnetic monopole \cite{jack,plyu,ikl}.\\

\noindent{\it 6.4 Isometries}
\vspace{0.2cm}

One may ask what is the characteristic feature of the subclass
of $q^+$ actions with which
we end up after substituting the composite superfield
${\cal V}^{ik}$ (or ${\cal V}^{++}$) into
a general $V^{++}$ action. The answer is that such a subclass
is distinguished by its U(1)
symmetry down to which the constant vector $a_{ab}$ breaks the
Pauli-G\"ursey SU(2) and under which
\p{composite}, \p{compos2} and \p{compcompos} are manifestly invariant.
This U(1) is an isometry of
the relevant target-space metric and one-form potential. In the case of $n$
superfields $q^{+a}_B$
one can define $n$ composite superfields ${\cal V}_{ia\,A}$ via the recipe
\p{composite} with $n$
independent constant tensors $a_{ab}$. Substituting such composite superfields
into the generic
$V^{++}_B$ action gives rise to the subclass of $q^+$ actions which yield $n$
commuting U(1) isometries
of the rotational type.

Like in the case of $N{=}2, d=4$ hypermultiplets in harmonic
superspace \cite{book},
one can define on $q^{+a}$ also an isometry of the translational type,
\be
q^{+a}{}' = q^{+ a} + \lambda u^{+a}\, \quad \mbox{or} \quad q^{ia}{}'
= q^{ia} + \lambda \epsilon^{ai}\,.
\label{translisom}
\ee
The defining constraint \p{qconstr} is evidently invariant under
\p{translisom}. Projecting $q^{+a}$ on the harmonics via
\be
l^{++} \equiv q^{+a}u^+_a\, \quad \omega \equiv q^{+a}u^-_a\, ,
\ee
one finds that
\be
l^{++}{}' = l^{++}\,, \quad \omega{}' = \omega + \lambda\,,
\ee
so the $q^{+a}$ Lagrangians which respect this isometry
are characterized by their independence of $\omega$. Observing that
the defining $q^{+a}$
constraint \p{qconstr} implies
\be
D^{++} l^{++} = 0\,,
\ee
we conclude that $l^{++}$ is again a composite $({\bf 3,4,1})$ multiplet with
\be
l^{ik} = \delta^{(i}_af^{k)\,a}\,, \quad \psi^i = \delta^i_a\chi^a\,,
\quad f = \epsilon_{ai}\dot{f}^{ia}\,.
\ee
This linear type of correspondence between the multiplets
$({\bf 4, 4, 0})$ and $({\bf 3,4,1})$ is
just the one discussed in \cite{pt}. General $q^{+}$ actions possessing this
type of isometry
can be obtained by substituting $l^{++}$ for $V^{++}$ into generic actions
for the latter,
discussed in Sect. 5.1. At the level of bosonic actions, this can be rephrased:
Imposing invariance under the isometry \p{translisom} forces the general
$f^{ia}$ actions \p{bosqgen},
\p{bosqA} to coincide with those for $v^{ik} = l^{ik}$, with the time derivative
of the fourth coordinate
$\epsilon_{ai}\dot{f}^{ia}$ mimicking the auxiliary field $F$. This agrees with
the reduction
procedure from the sigma models based on $({\bf 4, 4, 0})$ multiplets to those
with $({\bf 3,4,1})$ multiplets,
as it was  described in \cite{hp}. Obviously, this works for any number
of $q^{+a}_A$ multiplets.

Finally, we consider two examples.

Requiring $L(q)$ in \p{qgener1} to depend only on the symmetric combination
$q^{(ia)}$ and so be
invariant under \p{translisom} gives us the following particular case
of an ${\bf R}^4$ bosonic
sigma-model action \p{bosqgen}:
\be
\hat{S}_{(q)}^{red} = \int dt \;\Delta L({\bf f})\,
\left(\dot{\bf f}\cdot \dot{\bf f} + \sfrac{1}{2}\dot{f}\dot{f} \right)\,, \quad
\Delta = \partial^2/\partial {\bf f}\cdot \partial {\bf f }\,.
\ee
This should be compared with the general bosonic $v^{ik}$ action \p{bos1}.

As another example, let us consider the subclass of $q^{+a}$ superpotentials
\p{Supq} invariant under \p{translisom}:
\be
S_{(q)}^{1\ sp} = -\sfrac{i}{2}\int dud\zeta^{--} {\cal L}^{++}(q^+u^+, u)\,.
\ee
The corresponding one-form potential defined in \p{bosqA} takes the form
\be
{\cal A}_{ia}({\bf f})  = \int du\, u^-_iu^+_a
\frac{\partial {\cal L}^{++}}{\partial (f^{+}u^+)}\,.
\ee
Splitting it into the ${\bf R}^3$ and fourth components,
\be
{\cal A}_{(ia)}({\bf f}) = \int du\, u^-_{(i}u^+_{a)}
\frac{\partial {\cal L}^{++}}{\partial (f^{+}u^+)}\,, \quad
{\cal A}_4 ({\bf f}) = \epsilon^{ia}{\cal A}_{ia} = \int du\,
\frac{\partial {\cal L}^{++}}{\partial (f^{+}u^+)}\,,
\ee
we see that they coincide with the ${\bf R}^3$ one-form potential
and scalar half-potential
\p{VA} of the general superpotential for the superfield $V^{++}$ (with $v^{++} =
f^{ia}u^+_iu^+_a$). The constraints \p{monop} are recognized as the `static' monopole
ansatz for
the ${\bf R}^4$ self-duality equation \p{selfd1}. The general bosonic $f$/${\cal A}$
coupling \p{bosqA} for
this particular case takes the form
\be
\hat{S}_{(q)}^{1\ sp} = \int dt\left(\dot{\bf f}\cdot {\bf {\cal A}}
+ \sfrac12\dot{f} {\cal A}_4 \right)\,.
\ee
The SO(4) symmetry is clearly broken in this action. An ${\bf R}^4$ particle with
this type of
interaction  can be interpreted as a dyonic particle: the 4th component of its position
field is coupled to a static electric
potential while the ${\bf R}^3$ component sees a static magnetic field. A particular SO(3)
invariant
example describes a coupling to the Coulomb potential \p{qul} and the Dirac monopole
potential \p{confVect}.
%{\bf [ In our previous paper the electric field was not Coulomb but $\sim 1/r^3$ ]}\newline
Note that the corresponding $N{=}4$ supersymmetric Lagrangian is obtained by performing
the substitution
$\hat{V}^{++} \rightarrow (q^{+a}u^{+}_a) - c^{++}$ in the superconformally invariant
action \p{confSpot}.
The resulting $q^{+a}$ action, however, is not superconformally invariant, since the
transformation law
of $l^{++} = q^{+a}u^+_a$ under $D(2,1;\alpha)$ is different from that of $V^{++}$,
\be
\delta' l^{++} = \Lambda\, l^{++} + \Lambda^{++}\,\omega \,.
\ee

Finally, we note that for the $q^{+a}$ actions the geometries of the sigma-model
part and those of
the superpotential part do not correlate with each other like in the $V^{++}$ case,
and so one can
impose invariance requirements on these pieces separately. Typically the full actions
then are not
obliged to respect any symmetry except for $N{=}4, d{=}1$ Poincar\'e supersymmetry.

\setcounter{equation}{0}
\section{Concluding remarks}
In this paper we have introduced the $N{=}4, d{=}1$ harmonic superspace as a new setting
for SQM models with $N{=}4$ supersymmetry. We have shown that the off-shell $({\bf 3,4,1})$
and $({\bf 4,4,0})$
multiplets, which were utilized earlier for $N{=}4$ SQM model building in the framework of
standard $N{=}4, d{=}1$
superspace, have a natural description as constrained analytic harmonic superfields
$V^{++}(\zeta, u)$ and $q^{+a}(\zeta, u)$, respectively. The analytic harmonic superspace
was shown to be closed under
the most general $N{=}4, d{=}1$ superconformal group $D(2,1;\alpha)$ at any value of the
parameter $\alpha$, and the realization of this supergroup on $V^{++}$ and $q^{+a}$ was
found.
We presented harmonic superspace actions for these superfields, both in the superconformal
and
in the generic cases, and demonstrated that the conditions on the
bosonic target-space metrics, scalar potentials and one-forms required by $N{=}4$ supersymmetry
are automatically
reproduced from this manifestly $N{=}4$ supersymmetric off-shell description.
The superpotential-type pieces of
the full action are given by integrals over the $(1+2|2)$-dimensional analytic subspace of
$N{=}4, d{=}1$
harmonic superspace, and it is the only possibility to write down such terms off shell
in a manifestly $N{=}4$ supersymmetric manner.

As one problem for future study, it is tempting to construct and examine SQM
models associated with more general superfields $q^{(+n)}(\zeta, u)$ subject to
$D^{++}q^{(+n)} = 0$
for $n{>}2$, which are suggested by the harmonic superspace approach.
% as the natural generalization of
%$V^{++}$ and $q^{+a}$,
Some other consrained analytic multiplets from $N{=}2, d{=}4$ harmonic superspace
\cite{book} also have $d{=1}$ analogs. For instance, one can define a nonlinear multiplet
$N^{++}(\zeta, u)$ by the constraint $D^{++}N^{++} + (N^{++})^2 = 0\,$. Although this
$N{=}4, d{=}1$
constraint has the same form as in the $N{=}2, d{=}4$ case, the dynamics
of $N^{++}$ in $N{=}4, d{=}1$ harmonic superspace should differ essentially,
in particular
due to the different dimension and harmonic U(1) charge of the $d{=}1$ analytic
superspace integration measure.

Other interesting problems one can try to attack within the harmonic superspace approach are
the setting up of new $N{=}8, d{=}1$ SQM models by combining several analytic $N{=}4$
multiplets
into an irreducible off-shell $N{=}8$ multiplet as well as the study of the relationship
between
superconformal $N{=}4, d{=}1$ models and superparticles on AdS$_2\times S^2$ as
a special case of the general AdS/CFT correspondence.

\section*{Acknowledgements}
The authors thank Sergey Krivonos for valuable discussions and remarks. E.I. is
grateful to Andrei Smilga for useful comments and illuminating discussions.
%O.L. would like to thank N.~Dragon for useful discussions on sect.~5.
The work of E.I. was supported in part by
an INTAS grant, project No 00-00254, DFG grant, project No 436
RUS 113/669, RFBR-DFG grant, project No 02-02-04002, RFBR-CNRS grant,
project No 01-02-22005, and a grant of the Heisenberg-Landau program.
He thanks the Institute for Theoretical Physics of the University of Hannover
for the warm hospitality extended to him during the course of this work.


\begin{thebibliography}{99}
\bibitem{review} R. de Lima Rodrigues, `The quantum mechanics SUSY algebra:
an introductory review',
{\tt hep-th/0205017}.
\bibitem{gauntl1} J.P. Gauntlett, N. Kim, J. Park, P. Yi, Phys. Rev. D 61 (2000) 125012,
{\tt hep-th/9912082}; \\
J.P. Gauntlett, Chan-ju Kim, Ki-Myeong Lee, P. Yi, Phys. Rev. D 63 (2001) 065020,
{\tt hep-th/0008031}.
\bibitem{nscm} P. Claus, M. Derix, R. Kallosh, J. Kumar, P.K. Townsend,
A. Van Proeyen,\\
Phys. Rev. Lett. 81 (1998) 4553, {\tt hep-th/9804177}.
\bibitem{will}N. Wyllard, J. Math. Phys. 41 (2000) 2826, {\tt hep-th/9910160}.
\bibitem{bgk} S. Bellucci, A. Galajinsky, S. Krivonos, ``New
many-body superconformal models as reductions of simple composite systems'',
{\tt hep-th/0304087}.
\bibitem{6} G.W. Gibbons, P.K. Townsend,
Phys. Lett. B 454 (1999) 187, {\tt hep-th/9812034}.
\bibitem{papa2}R.A. Coles, G. Papadopoulos, Class. Quant. Grav. 7 (1990) 427.
\bibitem{hull}  C.M. Hull, ``The geometry of supersymmetric quantum
mechanics'', {\tt hep-th/9910028}.
\bibitem{papa1}G. Papadopoulos, Class. Quant. Grav. 17 (2000) 3715,
{\tt hep-th/0002007}.
\bibitem{stro1}J. Michelson, A. Strominger, Commun. Math. Phys. 213 (2000) 1,
{\tt hep-th/9907191}; JHEP 9909 (1999) 005, {\tt hep-th/9908044}.
\bibitem{stro2}A. Maloney, M. Spradlin, A. Strominger, JHEP 0204 (2002) 003,
{\tt hep-th/9911001}.
\bibitem{ikl2} E. Ivanov, S. Krivonos, V. Leviant,
J. Phys. A: Math. Gen. 22 (1989) 4201.
\bibitem{ismi} E.A. Ivanov, A.V. Smilga, Phys. Lett. B 257 (1991) 79.
\bibitem{ikp} E.A. Ivanov, S.O. Krivonos, A.I. Pashnev,
Class. Quant. Grav. 8 (1991) 19.
\bibitem{7} J.A. de Azcarraga, J.M. Izquierdo, J.C. Perez Bueno,
P.K. Townsend,\\
Phys. Rev. D 59 (1999) 084015, {\tt hep-th/9810230}.
\bibitem{hp} S. Hellerman, J. Polchinski, ``Supersymmetric quantum mechanics
from light cone quantization'', In: Shifman, M.A. (ed.), ``The many faces of
the superworld'', {\tt hep-th/9908202}.
\bibitem{sorpa} E.E. Donets, A. Pashnev, V.O. Rivelles, D.P. Sorokin,
M. Tsulaia, \\
Phys. Lett. B 484 (2000) 337, {\tt hep-th/0004019}.
\bibitem{paN4} E.E. Donets, A. Pashnev, J. Juan Rosales, M.M. Tsulaia, \\
Phys. Rev. D 61 (2000) 043512, {\tt hep-th/9907224}.
\bibitem{ikl} E. Ivanov, S. Krivonos, O. Lechtenfeld, JHEP 0303 (2003) 014,
{\tt hep-th/0212303}.
\bibitem{gikos}A. Galperin, E. Ivanov, V. Ogievetsky, E. Sokatchev, Pis'ma ZhETF
40 (1984) 155 [JETP Lett. 40 (1984) 912];\\
A.S. Galperin, E.A. Ivanov, S. Kalitzin, V.I. Ogievetsky,
E.S. Sokatchev, \\
Class. Quant. Grav. 1 (1984) 469.
\bibitem{book} A.S. Galperin, E.A. Ivanov, V.I. Ogievetsky,
E.S. Sokatchev, ``Harmonic superspace'', Cambridge University Press, 2001,
306 p.
\bibitem{cromrit} M. De Crombrugghe, V. Rittenberg, Ann. Phys. 151 (1983) 99.
\bibitem{andr1} A.V. Smilga, ZhETF 91 (1986) 14 [Sov. Phys. JETP 64 (1986) 8].
\bibitem{andr2} A.V. Smilga, Nucl. Phys. B 291 (1987) 241.
\bibitem{bepa} V.P. Berezovoj, A.I. Pashnev, Class. Quant. Grav. 8 (1991) 2141.
%\bibitem{ikp}
\bibitem{IvSu} E. Ivanov, A. Sutulin,
Nucl. Phys. B 432 (1994) 246; B 483 (1997) 531E, {\tt hep-th/9404098}.
\bibitem{zu} B. Zupnik, Nucl. Phys. B 554 (1999)365; B 644 (2002) 405E,
{\tt hep-th/9902038}.
\bibitem{di} F. Delduc, E. Ivanov, Phys. Lett. B 309 (1993) 312,
{\tt hep-th/9301024}.
\bibitem{dik} F. Delduc, E. Ivanov, S. Krivonos,
J. Math. Phys. 37 (1996) 1356; 38 (1997) 1224E, {\tt hep-th/9510033}.
\bibitem{gio1} A. Galperin, E. Ivanov, V. Ogievetsky,
Yad. Fiz. 45 (1987) 245 [Sov. J. Nucl. Phys. \break 45 (1987) 157]; Phys. Scripta
T 15 (1987) 176.
\bibitem{gio2} A.S. Galperin, E.A. Ivanov, V.I. Ogievetsky,
Nucl. Phys. B 282 (1987) 74.
\bibitem{gps} G.W. Gibbons, G. Papadopoulos, K.S. Stelle,
Nucl. Phys. B 508 (1997) 623, {\tt hep-th/9706207}.
\bibitem{pt} A. Pashnev, F. Toppan, J. Math. Phys. 42 (2001) 5257,
{\tt hep-th/0010135}.
%\bibitem{hull} Hull
\bibitem{andrei} A.V. Smilga, JHEP 0204 (2002) 054, {\tt hep-th/0201048}.
\bibitem{dff} V. De Alfaro, S. Fubini, G. Furlan, Nuovo Cim. A 34 (1974) 569.
\bibitem{jack}R. Jackiw, Ann. Phys. (N.Y.) 129 (1980) 183.
\bibitem{fr} S. Fubini, E. Rabinovici, Nucl. Phys. B 245 (1984) 17.
\bibitem{bn} S. Bellucci, A. Nersessian, Phys. Rev. D 64 (2001) 021702,
{\tt hep-th/0101065}.
\bibitem{gh} G.W. Gibbons, S.W. Hawking, Phys. Lett. 78 B (1978) 430.
\bibitem{egh} T. Eguchi, P.B. Gilkey, A.J. Hanson, Phys. Rept.
66 (1980) 213.
\bibitem{hktoric} N.J. Hitchin, A. Karlhede, U. Lindstr\"om, M. Ro\v{c}ek, \\
Commun. Math. Phys. 108 (1987) 535; \\
H. Pedersen, Y.-S. Poon, Commun. Math. Phys. 117 (1988) 569.
\bibitem{gp} J. Gutowski, G. Papadopoulos, Phys. Rev. D 62 (2000) 064023,
{\tt hep-th/0002242}.
\bibitem{chs}C.G. Callan, Jr., J.A. Harvey, A. Strominger,
Nucl. Phys. B 359 (1991) 611; \\
B 367 (1991) 60.
\bibitem{ikol2} E. Ivanov, S. Krivonos, O. Lechtenfeld, work in progress.
\bibitem{plyu} M.S. Plyushchay, Nucl. Phys. B 589 (2000) 413,
{\tt hep-th/0004032};\\
Nucl. Phys. Proc. Suppl. 102 (2001) 248, {\tt hep-th/0103040}.

\end{thebibliography}
\end{document}